\documentclass[final,3p,times,authoryear]{elsarticle}

\usepackage{amssymb, bm}

\usepackage{caption}
\usepackage{subcaption}
\usepackage{graphicx}
\usepackage[dvips]{epsfig}
\usepackage{xtab}
\usepackage{booktabs}
\usepackage{lscape}
\usepackage{rotating}
\usepackage{natbib}
\usepackage{import} 
\usepackage[usenames, dvipsnames]{color}
\usepackage{csquotes}
\usepackage{float}
\usepackage{color, soul} 
\usepackage[normalem]{ulem} 
\usepackage{appendix}
\usepackage{longtable}
\usepackage{siunitx} 

\usepackage{multirow}

\usepackage{pifont}
\usepackage{etoolbox} 
\makeatletter
\patchcmd\@combinedblfloats{\box\@outputbox}{\unvbox\@outputbox}{}{%
	\errmessage{\noexpand\@combinedblfloats could not be patched}%
}%
\makeatother
\journal{Icarus}

\begin{document}

\begin{frontmatter}



\title{The Volatility Trend of Protosolar and Terrestrial Elemental Abundances}

\author[label1,label2]{Haiyang S. Wang\corref{cor1}}
\ead{haiyang.wang@anu.edu.au}
\author[label1,label2,label3]{Charles H. Lineweaver}
\author[label2,label3]{Trevor R. Ireland}

\address[label1]{Research School of Astronomy and Astrophysics, The Australian National University, Canberra, ACT 2611, Australia}
\address[label2]{Planetary Science Institute, The Australian National University, Canberra, ACT 2611, Australia}
\address[label3]{Research School of Earth Sciences, The Australian National University, Canberra, ACT 2601, Australia}
\cortext[cor1]{Corresponding author. Mailing address: Mt Stromlo Observatory (ANU Research School of Astronomy and Astrophysics), Cotter Road, Weston Creek, ACT 2611, Australia}


\begin{abstract}
We present new estimates of protosolar elemental abundances based on an improved combination of
solar photospheric abundances and CI chondritic abundances. 
These new estimates indicate CI chondrites and solar abundances are consistent for 60 elements. Our estimate of the protosolar \textquotedblleft metallicity" (i.e. mass fraction of metals, $Z$) is 1.40\%, which is consistent with a value of $Z$ that has been decreasing steadily over the past three decades from $\sim 1.9\%$.
We compare our new protosolar abundances with our recent estimates of bulk Earth composition (normalized to aluminium),
thereby quantifying the devolatilization in going from the solar nebula to the formation of the Earth.
The quantification yields a linear trend $\log(f) = \alpha\log(T_C) + \beta$, where $f$ is the Earth-to-Sun abundance ratio and $T_C$ is the 50\% condensation temperature of elements. 
The best fit coefficients are: $\alpha = 3.676\pm 0.142$ and $\beta = -11.556\pm 0.436$. 
The quantification of these parameters constrains models of devolatilization processes.
For example, the coefficients $\alpha$ and $\beta$ determine a critical devolatilization temperature for the Earth $T_{\mathrm{D}}(\mathrm{E}) = 1391 \pm 15$ K.
The terrestrial abundances of elements with  $T_{C} < T_{\mathrm{D}}(\mathrm{E})$ are depleted compared with solar abundances, 
whereas the terrestrial abundances of elements with $T_{C} > T_{\mathrm{D}}(\mathrm{E})$ are indistinguishable from solar abundances.
The abundances of noble gases and hydrogen are depleted more than a prediction based on the extrapolation of the best-fit volatility trend. 
The terrestrial abundance of Hg ($T_C$ = 252 K) appears anomalously high under the assumption that solar and CI chondrite Hg abundances are identical.
To resolve this anomaly, we propose that CI chondrites have been depleted in Hg relative to the Sun by a factor of $13\pm7$.
We use the best-fit volatility trend to derive the fractional distribution of carbon and oxygen between volatile and refractory components ($f_\mathrm{vol}$, $f_\mathrm{ref}$).
For carbon we find ($0.91\pm 0.08$, $0.09 \pm 0.08$); for oxygen we find ($0.80 \pm 0.04$, $0.20 \pm 0.04$). 
Our preliminary estimate gives CI chondrites a critical devolatilization temperature $T_{\mathrm{D}} (\mathrm{CI}) = 550^{+20}_{-100}$ K.  
\end{abstract}

\begin{keyword}
Elemental abundance \sep Volatility trend \sep Proto-Sun \sep CI chondrites \sep Earth


\end{keyword}

\end{frontmatter}



\section{Introduction}
\label{sec:intro}

To first order, Earth is a devolatilized piece of the solar nebula.
Similarly, rocky exoplanets are almost certainly devolatilized pieces of the stellar nebulae out of which they and their host stars formed.
If this is correct, we can estimate the chemical composition of rocky exoplanets by measuring the elemental
abundances of their host stars, and then applying a devolatilization algorithm. 
The main goal of this paper is to go beyond the usual comparison of the silicate Earth with CI chondrites. 
We do this by comparing the bulk elemental abundances of Earth and Sun, and thus calibrate this potentially universal 
process associated with the formation of terrestrial planets. 

Determining the chemical abundances of Earth and Sun is not straightforward.  
For Earth, the composition must be obtained by determining the chemical abundances in primitive mantle and core, and by determining the mass fractions of these respective reservoirs.  
Different geochemical models make different underlying assumptions about the behavior of specific elements.  
However, for the purposes of this work (and for any comparative analysis), justifiable determinations of the contributions of these differences to the systematic uncertainties in the elemental abundances are required. Thus, for our analysis here, we use the bulk Earth abundances and their uncertainties from \cite{Wang2018}. We extend our modeling of terrestrial abundances to a comparison with solar abundances.  This will help quantify the processes that led to the rocky planets of our solar system, as well as by extension, to rocky exoplanets.

For the Sun, chemical abundances of a large number of elements can be determined spectroscopically, specifically by observing characteristic absorption lines in the solar photosphere.  Such determinations require accurate models of the solar circulation and calibration based on radiative transfer calculations.  Such calculations typically result in abundances with relatively large uncertainties, compared to the precision available from laboratory geochemical analyses. 
CI chondrites are therefore frequently used as a proxy in the determination of the relative abundances of many refractory elements in the Sun. 
Nevertheless, the abundances of some elements in CI chondrites are not representative of the abundances in the Sun.  
These include the most highly volatile elements  (H, He and the other noble gases) as well as other elements with significant gas phase abundances (e.g. C, N, and O).

A characteristic feature of the comparison between solar abundances and terrestrial abundances is the depletion for bulk Earth of elements with moderately low condensation temperatures, i.e., between 500 K and 1400 K. The depletion is systematic: the lower the condensation temperature, the greater the depletion. A version of volatile depletion has been previously noted by comparing the Earth's primitive mantle (i.e. the silicate part of the Earth by excluding the core) to CI chondrites, and called the \textquotedblleft volatility trend" of silicate Earth \citep{Kargel1993, McDonough1995, Palme2003}. 

A full quantification of the volatility trend between bulk Earth and proto-Sun provides a framework for understanding the nature of devolatilization processes active on the precursors to Earth. It allows a statistically robust determination of whether elements lie on the trend or not. Systematic deviations of one or more elements could indicate multistage processing, or processing that is not strictly related to temperature. Eventually, an extension of this model could be used to estimate rocky exoplanet compositions based on the elemental composition of their host stars.

Throughout this paper the word \textquotedblleft devolatilization" simply indicates the observed depletion feature of volatiles in Earth compared to Sun, with no particular reference to whether the depletion of an element happened during the collapse of the solar nebula, during accretion within the proto-planetary disk or subsequently as a result of impacts \citep[cf.][]{Siebert2018, Hin2017, Norris2017, Albarede2009}.

\section{Protosolar elemental abundances}
\label{sec:solarabu}
\begin{figure*}[!ht]
	\begin{center}
		\includegraphics[trim=5.0cm 1.0cm 5.0cm 1.5cm, scale=0.7,angle=0]{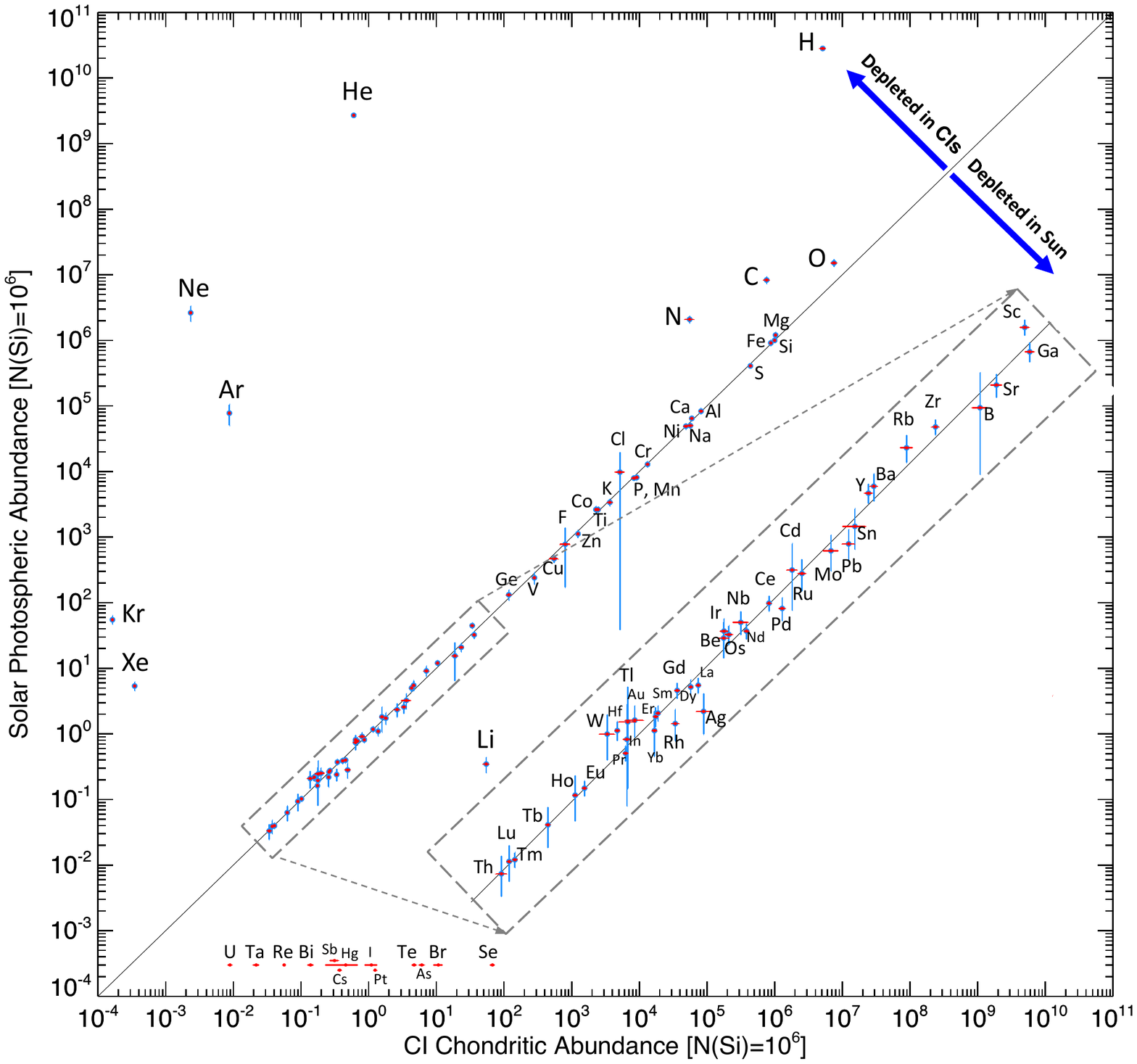} 
		\caption{Similarity of meteoritic and photospheric abundances.
			Abundances have been normalized to $10^6$ atoms of Si (Table \mbox{\ref{tab:sun}}, columns 6 and 10). 
			Points along the diagonal have identical abundances in both data sets. 
			Elemental abundances above the diagonal are depleted in CIs ( N, C, O, H and noble gases).
			Elemental abundances below the diagonal are depleted in the Sun (Li).  
			Photospheric abundances of 13 elements are unavailable.  They are set arbitrarily at $3\times10^{-4}$ and plotted on the left just above the x-axis. Uncertainties of CI chondritic abundances of noble gases are unavailable.}
		\label{fig:sun-1}
	\end{center}
\end{figure*}
\subsection{Meteoritic and photospheric elemental abundances}
\label{sec:solarabu1}
Calculation of protosolar elemental abundances requires assessment of two main sources of data: analytical data from the most primitive CI chondrite meteorites, and photospheric data based on spectroscopic observations from the Sun. Observations for many elements are available from both data sets, while some calculations can only be based on one source. 

Analyses of CI chondrites yield abundances of 83 stable elements -- many with high precision. Our analysis uses the CI chondritic abundances from the latest compilation by \cite{Palme2014a}, which are based on data from \cite{Lodders2009}, \cite{Barrat2012}, and \cite{Pourmand2012}. However, it is well known that not all these elemental abundances reflect protosolar abundances. Highly volatile elements, such as H, C, N, O, and the noble gases, are all significantly depleted in CI chondrites relative to solar abundances \citep{Anders1989, Lodders2003}.

Fewer elemental abundances can be inferred from spectroscopic observations of the solar photosphere. 
The determinations depend on the availability of spectral lines, the accuracy of atomic data, solar model atmospheres and spectral line formation calculations. 
For example, solar atmosphere spectral-line abundances are unavailable for 13 elements (As, Se, Br, Sb, Te, I, Cs, Ta, Re, Pt, Hg, Bi, and U). 
Lines of some other elements  (e.g. F, Cl, I and Tl) can only be identified in the spectra of sunspots because sunspot temperatures are lower than average photospheric temperatures \citep[][]{Asplund2009}. 
The abundance uncertainties of these elements are correspondingly larger. 

Due to lithium burning in the interior of the Sun and to mixing between the interior and the photosphere, lithium is depleted in the present-day solar photosphere. 
In the absence of available photospheric lines for the noble gases He, Ne, Ar, Kr, and Xe, the photospheric abundances of these elements are estimated through helioseismology, solar winds/corona/flares, and theoretical calculations \citep{Asplund2009}.

Our compilation of photospheric abundances (Table \mbox{\ref{tab:sun}}, column 3) are from \cite{Asplund2009}, \cite{Scott2015a, Scott2015b}, and \cite{Grevesse2015}.  These are based on a 
3D hydrodynamic solar model atmosphere and non-LTE (= non-Local Thermodynamic Equilibrium) spectral line calculations, in which statistical and systematic abundance uncertainties have been uniformly estimated. 
Disagreements on the solar photospheric abundance of oxygen from different 3D models of different groups \citep{Asplund2009, Caffau2015, Steffen2015} have been resolved by improved spectroscopic modeling as well as observational data \citep[see][]{Caffau2017, Amarsi2018}. These latest results are consistent with our adoption of the solar oxygen abundance from \cite{Asplund2009}. The most recent updates for 3D non-LTE abundances of Si, Fe, and Al in the Sun \citep{Amarsi2017, Lind2017, Nordlander2017} are also in excellent agreement with our adopted values from \cite{Scott2015a, Scott2015b}.

The comparison between the solar photospheric abundances and CI chondritic abundances is shown in Fig. \ref{fig:sun-1}. 
The abundances are nearly identical with some exceptions:
i) the 13 elements in the lower left do not have photospheric observations; 
ii) the Li abundance is depleted in the photosphere but is well-preserved in CI chondrites; 
iii) 9 highly volatile elements are moderately or severely depleted in CI chondrites, but their abundances can be determined through indirect observations and theoretical modeling of solar spectra. 
In general, the precision of meteoritic data (red) is higher than that of the photospheric data (blue). 

\subsection{Methods Used to Combine Photospheric and Meteoritic Abundances}
\label{sec:solarabu2}

\cite{Lodders2009} compiled and combined meteoritic and photospheric data to estimate protosolar abundances. Such a compilation can also be found in earlier studies \citep[e.g.][]{Anders1989, Grevesse1998, Lodders2003}.
The methodology adopted in these previous works generally entails using the high-precision meteoritic data and adding in photospheric data when CI data is likely compromised. \cite{Lodders2009} used an average of CI and solar data when solar data was consistent with CI data and was sufficiently precise.
We update this widely-cited compilation 
by combining the latest photospheric-based solar abundances and CI chondritic abundances 
using the following method (see column 12 of Table \ref{tab:sun} as well as Table \ref{tab:sm}). 
(1) The 9 elements labeled \textquoteleft p'- are depleted in meteorites and therefore only the photospheric-based solar abundances are used.  
(2) The 13 elements labeled \textquoteleft m'- have not had their photospheric abundances measured. In addition, Li is included in this group because its
current photospheric abundance is depleted. 
(3) The abundance of an element labeled \textquoteleft a' is the weighted average of its photospheric-based solar abundance and its meteoritic abundance. 
We place 60 elements in this category, while \cite{Lodders2009} places 26, attesting to the improvement in photospheric data quality.
Also, we use diffusion corrections to convert photospheric abundances to bulk solar abundances before (rather than after) they are normalized and combined with meteoritic data. This is because diffusion and settling issues affect photospheric abundances, not meteoritic abundances.
For more details see \mbox{\ref{sec:app1}}. For the conversion of meteoritic abundances from a silicon to a hydrogen normalization see \ref{sec:app2}.

Compared with the method of \mbox{\cite{Lodders2009}}, our method of combining meteoritic and photospheric data rejects less of the photospheric data and places more emphasis on averaging the two independent data sets.
\cite{Lodders2009} has placed 48 elements in the second (meteorite-only) group either because 
the photospheric and meteoritic abundances were inconsistent, 
or because the uncertainty on the photospheric abundance was substantially larger than the uncertainty on the meteoritic abundance.
Including photospheric abundances based partly on their similarity with the meteoritic evidence is not advisable \citep{Scott2015a}, since both meteoritic and photospheric data are only proxies of the bulk protosolar abundances.
Both have limitations. 
The most primitive CI chondrite group comprises only five meteorites for a total mass of only 17 kg.  Of these, Orgeuil is the largest with around 10 kg recovered, and this is the meteorite with the largest number of independent measurements \citep{Barrat2012}.  As such, the abundances from CI chondrites are based on an extremely small subsample of meteorites. 
On the other hand, photospheric measurments are limited by the precision of the extracted spectral lines, the accuracy of the solar models being applied, and the accuracy of the diffusion corrections performed. 
The agreement between CI chondrite abundances and solar photospheric abundances therefore needs to be reconfirmed in each analysis combining the two data sets.  
We therefore, where possible, have used weighted means to take advantage of the information in both sources.

\subsection{Protosolar elemental abundance results}
\label{sec:solarabu3}

\begin{figure*}[!ht]
	\begin{center}
		\includegraphics[trim=1.2cm 5.5cm 2.0cm 5.0cm, scale=0.85,angle=90]{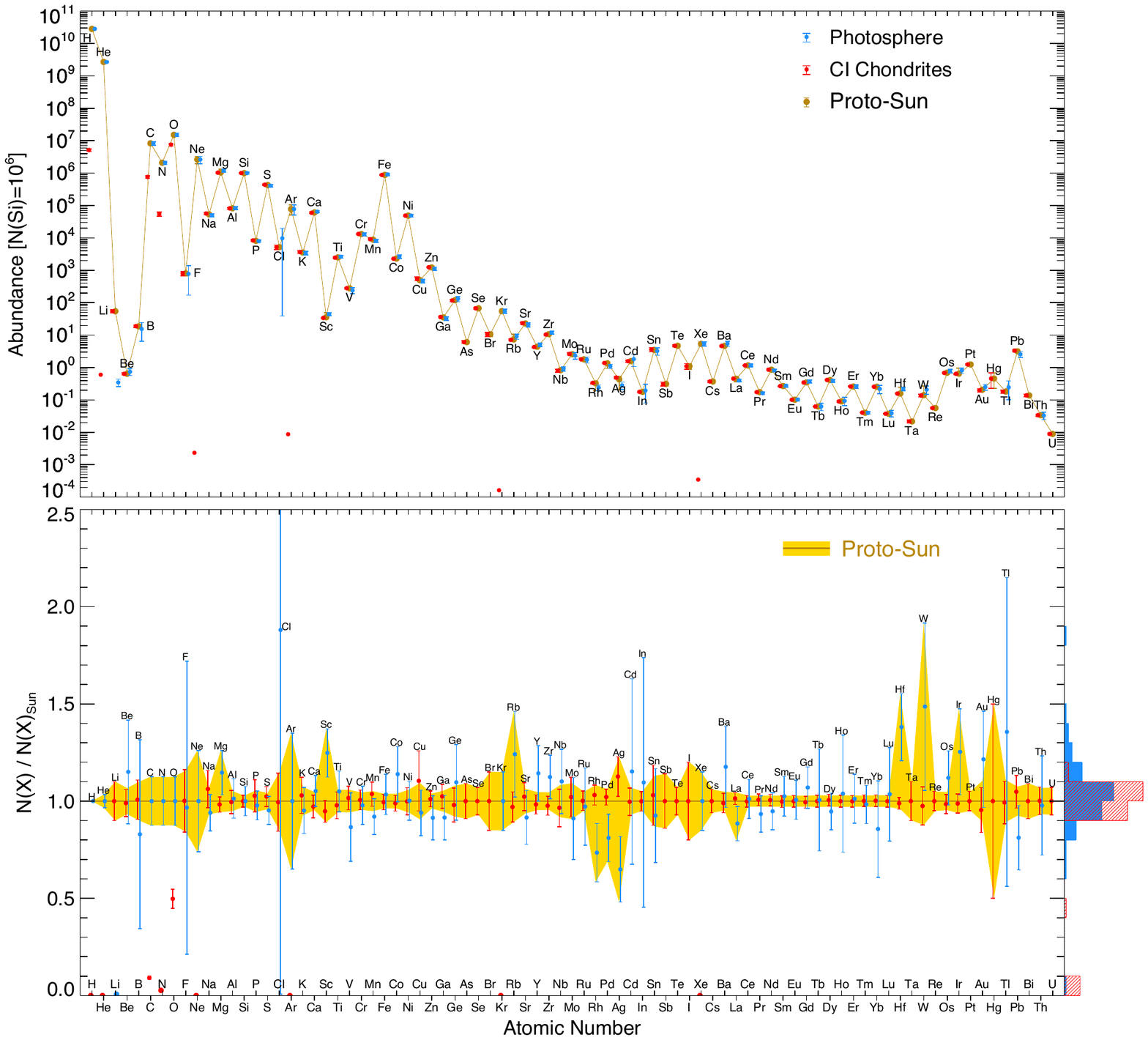}
		\caption{Our estimates of protosolar elemental abundances (Table \ref{tab:sun}, column 13) compared with the source data: the photospheric-based solar abundances and the meteoritic abundances (columns 6 and 10 of Table \ref{tab:sun}). In the lower panel, abundances have been normalized to our protosolar abundances. The upper limit of chlorine is off the plot at 3.76. The blue and red histograms along the right side of the lower panel show the statistical distributions of the photospheric and meteoritic data respectively. See discussion in Section \ref{sec:solarabu3} for a non-parametric Mann-Whitney U test on the data that produce the two histograms.}	
		\label{fig:sun-2}
	\end{center}
\end{figure*}

Our estimates of protosolar elemental abundances, compared with both the photospheric-based and meteoritic-based abundances, are listed in Table \ref{tab:sun} and 
plotted in Fig. \ref{fig:sun-2}. 
Fig. \mbox{\ref{fig:sun-2}} illustrates our selection methods and how the two data sets were combined to produce protosolar abundances.
The resultant protosolar nebula abundance of Li is determined solely from meteorite data, whereas those of C, N and O are based solely on the photospheric-based solar abundances. 
The weighted mean of a meteoritic abundance with a small error bar, and a photometric abundance with a large error bar, produces a result very similar to
the meteoritic abundance (e.g. F and Cl). 
When the uncertainties from the two data sets do not overlap we assign the upper and lower limits of the two data points, as conservative error bars on the weighted mean (e.g. Rh, Pd and Ag). 

Along the right side of the lower panel of Fig. \ref{fig:sun-2}, the blue and red histograms show the distributions of the photospheric and meteoritic residuals respectively. Both are centered at 1 with the blue histogram having a larger variance. To verify that the residuals shown in the red and blue histograms are consistent with coming from the same distribution, we apply a non-parametric Mann-Whitney U test to the residuals after excluding the apparent outliers (i.e., the red and blue dots along the x-axis and CI oxygen). We find the test statistic $Z=0.42$. The probability ($p-$value) of obtaining a $Z$ value $\ge$ 0.42 is 0.337. Since $p=0.337$ is greater than the fiducial 0.05 significance level, the two histograms are consistent with sampling a distribution with the same median value. A comparison between our estimates and the estimates of \cite{Lodders2009} for protosolar elemental abundances can be found in Fig. \ref{fig:suncompare}.

\begin{figure*}[!htbp]
\begin{center}
	\includegraphics[trim=1.2cm 5.5cm 2.0cm 5.0cm, scale=0.6,angle=0]{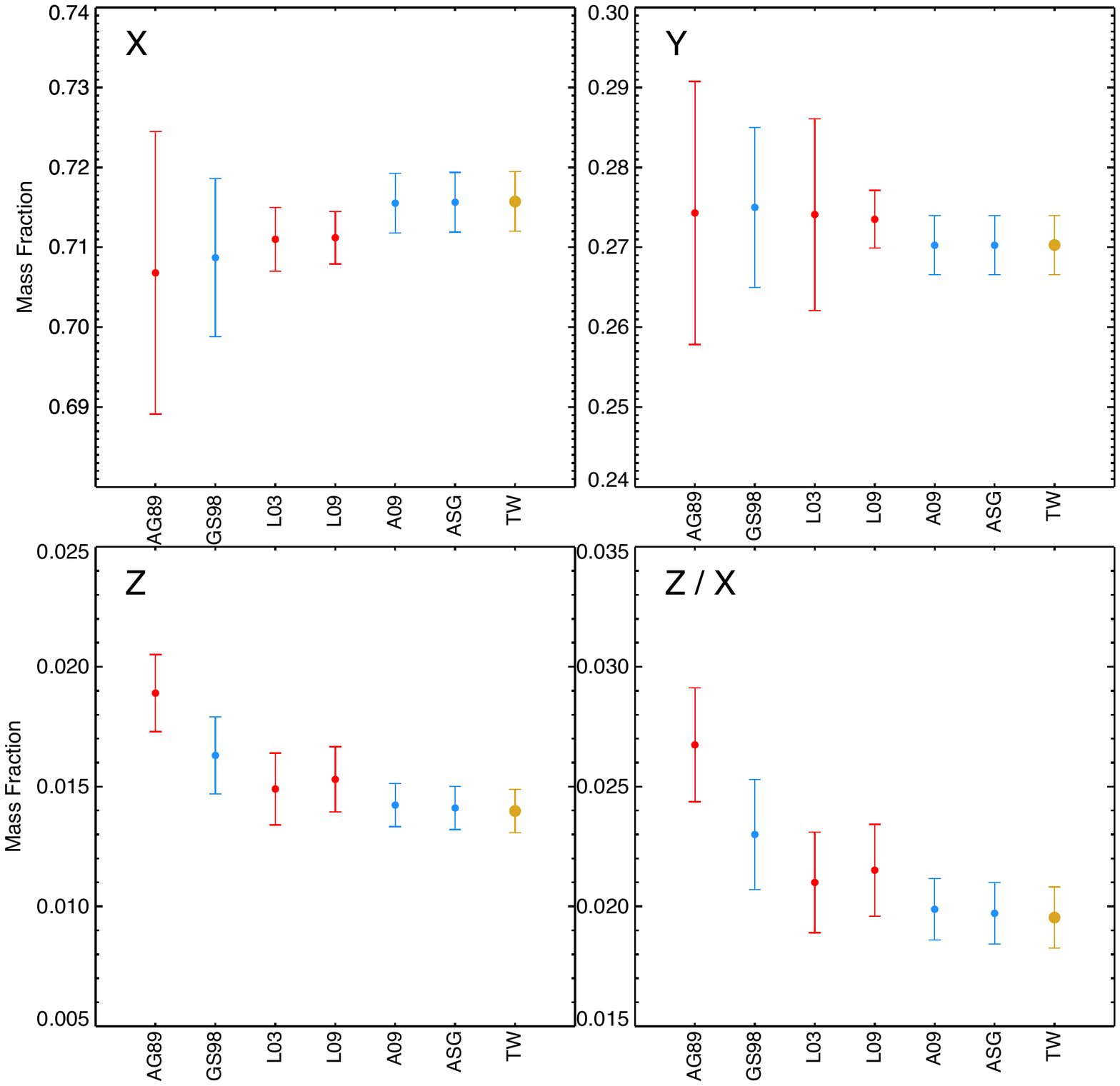}  
	\caption{The mass fractions of H ($X$), He ($Y$) and metals ($Z$) for protosolar elemental abundances in the literature and this work (Table \ref{tab:solarfrac}). AG89: \cite{Anders1989}, GS98: \cite{Grevesse1998}, L03: \cite{Lodders2003}, L09: \cite{Lodders2009}, A09:\cite{Asplund2009}, ASG: The bulk solar abundance compiled from \cite{Asplund2009}, \cite{Scott2015a, Scott2015b} and \cite{Grevesse2015}, TW: This work. For clarity, literature results based mostly on meteoritic data are red. Those based mostly on photospheric data are blue. The weighing used for our results (yellow) is described in Table \ref{tab:sm}.}
	\label{fig:massfra}
\end{center}
\end{figure*}

From an astronomical perspective, the elemental abundances of a star can be simplified to include only the mass fraction abundances of hydrogen (X), helium (Y) and \textquotedblleft metallicity" (the sum of everything else) (Z), 
where $X + Y + Z = 1$. 
For the bulk protosolar elemental abundances reported here (column 14 of Table \mbox{\ref{tab:sun}}), we estimate $X = 0.7157\pm0.0037$, $Y = 0.2703\pm0.0037$, and $Z = 0.0140\pm0.0009$.
For comparison, we compile and compute protosolar mass fractions reported in the literature over the past three decades (Table \ref{tab:solarfrac}).
There has been a steady decrease in $Z$ from $1.89\%$ in \cite{Anders1989} to $1.40\%$ in the present work (see Fig. \mbox{\ref{fig:massfra}}). This decrease is attributed mostly to the improvement in photospheric data quality, and is also reflected in the numerical factor $\Delta$ used to convert Si-normalized meteoritic abundances to H-normalized abundances -- for astronomical comparison in dex -- as discussed in \mbox{\ref{sec:app2}}.



\begin{landscape}
	{\scriptsize
		\begin{longtable}{ll  llll  lllll  lll} 
			\label{tab:sun}\\
			\caption[Estimates of protosolar abundances from photospheric and CI chondritic data]{Estimates of protosolar abundances from photospheric and CI chondritic data $\dagger$.}\\
			\multicolumn{1}{c}{1} & \multicolumn{1}{c}{2} & \multicolumn{1}{c}{3} & \multicolumn{1}{c}{4} & \multicolumn{1}{c}{5} & \multicolumn{1}{c}{6} & \multicolumn{1}{c}{7} & \multicolumn{1}{c}{8} & \multicolumn{1}{c}{9} & \multicolumn{1}{c}{10} & \multicolumn{1}{c}{11} & \multicolumn{1}{c}{12} & \multicolumn{1}{c}{13} & \multicolumn{1}{c}{14} \\
			\toprule 
			\multicolumn{2}{c}{} & \multicolumn{4}{c}{\bf Photosphere} & \multicolumn{5}{c}{\bf CI-Chondrites} & \multicolumn{3}{c}{\bf Proto-Sun} \\
			\cmidrule(lr){3-6} \cmidrule(lr){7-11} \cmidrule(lr){12-14} 
			Z  & Elm & $A(\textup{X})_p$$^a$ & Ref.$^b$ & $A(\textup{X})_{p,0}$ $^c$ & $N(\textup{X})_{p,0}$ $^d$ & $X_{ppm}$ $^e$ & $\sigma_{X}$ [\%] & Ref.$^f$ & $N(\textup{X})_{m}$ $^g$ & $A(\textup{X})_m$ $^h$ & SM$^i$ & $N(\textup{X})_{0}$ $^j$ & $A(\textup{X})_0$ $^k$ \\
			\hline  
			\endfirsthead
			\multicolumn{1}{c}{1} & \multicolumn{1}{c}{2} & \multicolumn{1}{c}{3} & \multicolumn{1}{c}{4} & \multicolumn{1}{c}{5} & \multicolumn{1}{c}{6} & \multicolumn{1}{c}{7} & \multicolumn{1}{c}{8} & \multicolumn{1}{c}{9} & \multicolumn{1}{c}{10} & \multicolumn{1}{c}{11} & \multicolumn{1}{c}{12} & \multicolumn{1}{c}{13} & \multicolumn{1}{c}{14} \\
			\toprule 
			\multicolumn{2}{c}{} & \multicolumn{4}{c}{\bf Photosphere} & \multicolumn{5}{c}{\bf CI-Chondrites} & \multicolumn{3}{c}{\bf Proto-Sun} \\
			\cmidrule(lr){3-6} \cmidrule(lr){7-11} \cmidrule(lr){12-14} 
			Z  & Elm & $A(\textup{X})_p$$^a$ & Ref.$^b$ & $A(\textup{X})_{p,0}$ $^c$ & $N(\textup{X})_{p,0}$ $^d$ & $X_{ppm}$ $^e$ & $\sigma_{X}$ [\%] & Ref.$^f$ & $N(\textup{X})_{m}$ $^g$ & $A(\textup{X})_m$ $^h$ & SM$^i$ & $N(\textup{X})_{0}$ $^j$ & $A(\textup{X})_0$ $^k$ \\
			
			\hline  
			\endhead
			\hline
			\multicolumn{14}{r}{\textit{Continued on next page}} \\
			\endfoot		
			\bottomrule	
			\multicolumn{14}{p{19cm}}{\tiny $^\dagger$ A machine-readable version of this table is available online.} \\
			\multicolumn{14}{p{19cm}}{\tiny $^a$ Solar photospheric abundances (in dex).} \\
			\multicolumn{14}{p{19cm}}{\tiny $^b$ References for the photosphere: A09-Columns 3 and 7 of Table 1 in \cite{Asplund2009}, S15a- Column 3 of Table 5 in \cite{Scott2015a}, S15b- Column 3 of Table 6 in \cite{Scott2015b}, G15- Column 3 of Table 5 in \cite{Grevesse2015}.}\\ 
			\multicolumn{14}{p{19cm}}{\tiny $^c$ Bulk solar abundances (in dex) by doing overall diffusion corrections on photospheric abundances ($A(\mathrm{X})_{p}$ in column 3) through Eq. \ref{eq:A2} and \ref{eq:A3}.} \\
			\multicolumn{14}{p{19cm}}{\tiny $^d$ Bulk solar abundances per $10^6$ atoms of Si, converted from $A(\mathrm{X})_{p,0}$ (column 5) by Eq. \ref{eq:A4}. Uncertainties are calculated using Eq. \ref{eq:A5}.} \\	
			\multicolumn{14}{p{19cm}}{\tiny $^e$ Elemental abundances in ppm by mass.} \\		
			\multicolumn{14}{p{19cm}}{\tiny $^f$ References for meteorites: L09- Columns 3 and 4 of Table 3 in \cite{Lodders2009}, P14- Columns 3 and 6 of Table 3 in \cite{Palme2014a}. \cite{Lodders2009} and \cite{Palme2014a} have used average ratios of chondritic meteorites of refractory lithophile and refractory sideroophile elements to improve the meteoritic CI-data.} \\
			\multicolumn{14}{p{19cm}}{\tiny $^g$ Meteoritic abundances per $10^6$ atoms of Si, converted from the elemental abundances in ppm by mass ($\mathrm{X_{ppm}}$ and $\sigma_{X}$ in columns 7 and 8) through Eqs. \ref{eq:A6} \& \ref{eq:A7}.} \\
			\multicolumn{14}{p{19cm}}{\tiny $^h$ Meteoritic abundances (in dex) converted by Eq. \ref{eq:B1}. using $\Delta = 1.554$. See \ref{sec:app2} for more details.} \\
			\multicolumn{14}{p{19cm}}{\tiny $^i$ SM- Selection methods. See Table \ref{tab:sm}.} \\
			\multicolumn{14}{p{19cm}}{\tiny $^j$ Protosolar abundance per $10^6$ atoms of Si, resulted from combining $N(\mathrm{X})_{p,0}$ (column 6) and $N(\mathrm{X})_{m}$ (column 10) through Eqs. \ref{eq:A8} \&  \ref{eq:A9}.}\\ 
			\multicolumn{14}{p{19cm}}{\tiny $^k$ Protosolar abundances (in dex), converted from $N(\mathrm{X_0}$) (column 13) by Eq. \ref{eq:A1}. Uncertainties are from solving Eq. \ref{eq:A5} for $\sigma_{A}$ using $\sigma_{N_{0}}$ instead of $\sigma_{N_{p,0}}$.}\\
			\multicolumn{14}{p{19cm}}{\tiny $^l$ This meteoritic Hg abundance does not reflect the factor of 13 $\pm$ 7 increase discussed in Section \ref{sec:discu2c}. Including this increase yields $5.954\pm4.375$ for $N(\mathrm{Hg})_0$ and $2.325\pm0.239$ for $A(\mathrm{Hg})_0$.} \\
			\endlastfoot	
				1	&	 H	&	12.00			&	A09	&	12.00			&	2.82			$  \times$	$10^{10}$	&	19700	&	10	&	L09/P14	&	5.13	$\pm$	0.51	$  \times$	$10^{6}$	&	8.26	$\pm$	0.04	&	p	&		2.82						$  \times$	$10^{10}$		&	\bf{	12.00						}	\\
2	&	 He	&	10.93	$\pm$	0.01	&	A09	&	10.98	$\pm$	0.01	&	2.69	$\pm$	0.09	$  \times$	$10^{9}$	&	9.17$\times10^{-3}$	&	-	&	L09/P14	&	0.60					&	1.33			&	p	&		2.69	$\pm$	0.09				$  \times$	$10^{9}$		&	\bf{	10.98	$\pm$	0.01				}	\\
3	&	  Li	&	1.05	$\pm$	0.10	&	A09	&	1.09	$\pm$	0.10	&	0.347	$\pm$	0.090			&	1.45	&	10	&	P14	&	54.6	$\pm$	5.5			&	3.29	$\pm$	0.04	&	m	&		54.6	$\pm$	5.5							&	\bf{	3.29	$\pm$	0.04				}	\\
4	&	  Be	&	1.38	$\pm$	0.09	&	A09	&	1.42	$\pm$	0.09	&	0.741	$\pm$	0.172			&	0.0219	&	7	&	P14	&	0.638	$\pm$	0.045			&	1.36	$\pm$	0.03	&	a	&		0.644	$\pm$	0.043							&	\bf{	1.36	$\pm$	0.03				}	\\
5	&	  B	&	2.70	$\pm$	0.20	&	A09	&	2.74	$\pm$	0.20	&	15.5	$\pm$	9.1			&	0.775	&	10	&	L09/P14	&	18.8	$\pm$	1.9			&	2.83	$\pm$	0.04	&	a	&		18.7	$\pm$	1.8							&	\bf{	2.82	$\pm$	0.04				}	\\
6	&	 C	&	8.43	$\pm$	0.05	&	A09	&	8.47	$\pm$	0.05	&	8.32	$\pm$	1.04	$  \times$	$10^{6}$	&	34800	&	10	&	L09/P14	&	7.60	$\pm$	0.76	$  \times$	$10^{5}$	&	7.43	$\pm$	0.04	&	p	&		8.32	$\pm$	1.04				$  \times$	$10^{6}$		&	\bf{	8.47	$\pm$	0.05				}	\\
7	&	 N	&	7.83	$\pm$	0.05	&	A09	&	7.87	$\pm$	0.05	&	2.09	$\pm$	0.26	$  \times$	$10^{6}$	&	2950	&	15	&	L09/P14	&	5.53	$\pm$	0.83	$  \times$	$10^{4}$	&	6.30	$\pm$	0.06	&	p	&		2.09	$\pm$	0.26				$  \times$	$10^{6}$		&	\bf{	7.87	$\pm$	0.05				}	\\
8	&	 O	&	8.69	$\pm$	0.05	&	A09	&	8.73	$\pm$	0.05	&	1.51	$\pm$	0.19	$  \times$	$10^{7}$	&	4.59$\times10^{5}$	&	10	&	L09/P14	&	7.53	$\pm$	0.75	$  \times$	$10^{6}$	&	8.43	$\pm$	0.04	&	p	&		1.51	$\pm$	0.19				$  \times$	$10^{7}$		&	\bf{	8.73	$\pm$	0.05				}	\\
9	&	  F	&	4.40	$\pm$	0.25	&	A09	&	4.44	$\pm$	0.25	&	776	$\pm$	605			&	58.2	&	16	&	P14	&	804	$\pm$	129			&	4.46	$\pm$	0.06	&	a	&		803	$\pm$	126							&	\bf{	4.45	$\pm$	0.06				}	\\
10	&	 Ne	&	7.93	$\pm$	0.10	&	A09	&	7.97	$\pm$	0.10	&	2.63	$\pm$	0.68	$  \times$	$10^{6}$	&	1.8$\times10^{-4}$	&	-	&	L09/P14	&	2.34			$  \times$	$10^{-3}$	&	-1.08			&	p	&		2.63	$\pm$	0.68				$  \times$	$10^{6}$		&	\bf{	7.97	$\pm$	0.10				}	\\
11	&	  Na	&	6.21	$\pm$	0.04	&	S15b	&	6.25	$\pm$	0.04	&	5.01	$\pm$	0.50	$  \times$	$10^{4}$	&	4962	&	9	&	P14	&	5.67	$\pm$	0.51	$  \times$	$10^{4}$	&	6.31	$\pm$	0.04	&	a	&		5.33	$\pm$	0.36				$  \times$	$10^{4}$		&	\bf{	6.28	$\pm$	0.03				}	\\
12	&	  Mg	&	7.59	$\pm$	0.04	&	S15b	&	7.63	$\pm$	0.04	&	1.20	$\pm$	0.12	$  \times$	$10^{6}$	&	95400	&	4	&	P14	&	1.03	$\pm$	0.04	$  \times$	$10^{6}$	&	7.57	$\pm$	0.02	&	a$^*$	&		1.05	$^{+	0.27	}_{-	0.06	}$	$  \times$	$10^{6}$		&	\bf{	7.57	$^{+	0.10	}_{-	0.03	}$	}	\\
13	&	  Al	&	6.43	$\pm$	0.04	&	S15b	&	6.47	$\pm$	0.04	&	8.32	$\pm$	0.83	$  \times$	$10^{4}$	&	8400	&	6	&	P14	&	8.17	$\pm$	0.49	$  \times$	$10^{4}$	&	6.47	$\pm$	0.03	&	a	&		8.21	$\pm$	0.42				$  \times$	$10^{4}$		&	\bf{	6.46	$\pm$	0.02				}	\\
14	&	  Si	&	7.51	$\pm$	0.03	&	S15b	&	7.55	$\pm$	0.03	&	1.00	$\pm$	0.08	$  \times$	$10^{6}$	&	1.07$\times10^{5}$	&	3	&	L09/P14	&	1.00	$\pm$	0.03	$  \times$	$10^{6}$	&	7.55	$\pm$	0.01	&	a	&		1.00	$\pm$	0.03				$  \times$	$10^{6}$		&	\bf{	7.55	$\pm$	0.01				}	\\
15	&	  P	&	5.41	$\pm$	0.03	&	S15b	&	5.45	$\pm$	0.03	&	7943	$\pm$	600			&	985	&	8	&	P14	&	8347	$\pm$	668			&	5.48	$\pm$	0.03	&	a	&		8124	$\pm$	446							&	\bf{	5.46	$\pm$	0.02				}	\\
16	&	  S	&	7.12	$\pm$	0.03	&	S15b	&	7.16	$\pm$	0.03	&	4.07	$\pm$	0.31	$  \times$	$10^{5}$	&	53500	&	5	&	L09/P14	&	4.38	$\pm$	0.22	$  \times$	$10^{5}$	&	7.20	$\pm$	0.02	&	a	&		4.28	$\pm$	0.18				$  \times$	$10^{5}$		&	\bf{	7.18	$\pm$	0.02				}	\\
17	&	Cl	&	5.50	$\pm$	0.30	&	S15b	&	5.54	$\pm$	0.30	&	9772	$\pm$	9734			&	698	&	15	&	L09/P14	&	5168	$\pm$	775			&	5.27	$\pm$	0.06	&	a	&		5197	$\pm$	773							&	\bf{	5.27	$\pm$	0.06				}	\\
18	&	Ar	&	6.40	$\pm$	0.13	&	S15b	&	6.44	$\pm$	0.13	&	7.76	$\pm$	2.72	$  \times$	$10^{4}$	&	1.33$\times10^{-3}$	&	-	&	L09/P14	&	8.74			$  \times$	$10^{-3}$	&	-0.50			&	p	&		7.76	$\pm$	2.72				$  \times$	$10^{4}$		&	\bf{	6.44	$\pm$	0.13				}	\\
19	&	K	&	5.04	$\pm$	0.05	&	S15b	&	5.08	$\pm$	0.05	&	3388	$\pm$	422			&	546	&	9	&	P14	&	3665	$\pm$	330			&	5.12	$\pm$	0.04	&	a	&		3560	$\pm$	260							&	\bf{	5.10	$\pm$	0.03				}	\\
20	&	Ca	&	6.32	$\pm$	0.03	&	S15b	&	6.36	$\pm$	0.03	&	6.46	$\pm$	0.49	$  \times$	$10^{4}$	&	9110	&	6	&	P14	&	5.97	$\pm$	0.36	$  \times$	$10^{4}$	&	6.33	$\pm$	0.03	&	a	&		6.14	$\pm$	0.29				$  \times$	$10^{4}$		&	\bf{	6.34	$\pm$	0.02				}	\\
21	&	Sc	&	3.16	$\pm$	0.04	&	S15a	&	3.20	$\pm$	0.04	&	44.7	$\pm$	4.4			&	5.81	&	6	&	P14	&	33.9	$\pm$	2.0			&	3.08	$\pm$	0.03	&	a$^*$	&		35.8	$^{+	13.3	}_{-	3.9	}$				&	\bf{	3.10	$^{+	0.14	}_{-	0.05	}$	}	\\
22	&	Ti	&	4.93	$\pm$	0.04	&	S15a	&	4.97	$\pm$	0.04	&	2630	$\pm$	262			&	447	&	7	&	P14	&	2451	$\pm$	172			&	4.94	$\pm$	0.03	&	a	&		2505	$\pm$	144							&	\bf{	4.95	$\pm$	0.02				}	\\
23	&	V	&	3.89	$\pm$	0.08	&	S15a	&	3.93	$\pm$	0.08	&	240	$\pm$	49			&	54.6	&	6	&	P14	&	281	$\pm$	17			&	4.00	$\pm$	0.03	&	a	&		277	$\pm$	16							&	\bf{	3.99	$\pm$	0.02				}	\\
24	&	Cr	&	5.62	$\pm$	0.04	&	S15a	&	5.66	$\pm$	0.04	&	1.29	$\pm$	0.13	$  \times$	$10^{4}$	&	2623	&	5	&	P14	&	1.32	$\pm$	0.07	$  \times$	$10^{4}$	&	5.68	$\pm$	0.02	&	a	&		1.32	$\pm$	0.06				$  \times$	$10^{4}$		&	\bf{	5.67	$\pm$	0.02				}	\\
25	&	Mn	&	5.42	$\pm$	0.04	&	S15a	&	5.46	$\pm$	0.04	&	8128	$\pm$	810			&	1916	&	6	&	P14	&	9154	$\pm$	549			&	5.52	$\pm$	0.03	&	a	&		8831	$\pm$	455							&	\bf{	5.50	$\pm$	0.02				}	\\
26	&	Fe	&	7.47	$\pm$	0.04	&	S15a	&	7.51	$\pm$	0.04	&	9.12	$\pm$	0.91	$  \times$	$10^{5}$	&	1.87$\times10^{5}$	&	4	&	P14	&	8.77	$\pm$	0.35	$  \times$	$10^{5}$	&	7.50	$\pm$	0.02	&	a	&		8.82	$\pm$	0.33				$  \times$	$10^{5}$		&	\bf{	7.50	$\pm$	0.02				}	\\
27	&	Co	&	4.93	$\pm$	0.05	&	S15a	&	4.97	$\pm$	0.05	&	2630	$\pm$	328			&	513	&	4	&	P14	&	2285	$\pm$	91			&	4.91	$\pm$	0.02	&	a	&		2310	$\pm$	88							&	\bf{	4.91	$\pm$	0.02				}	\\
28	&	Ni	&	6.20	$\pm$	0.04	&	S15a	&	6.24	$\pm$	0.04	&	4.90	$\pm$	0.49	$  \times$	$10^{4}$	&	10910	&	7	&	P14	&	4.88	$\pm$	0.34	$  \times$	$10^{4}$	&	6.24	$\pm$	0.03	&	a	&		4.89	$\pm$	0.28				$  \times$	$10^{4}$		&	\bf{	6.24	$\pm$	0.02				}	\\
29	&	Cu	&	4.18	$\pm$	0.05	&	G15	&	4.22	$\pm$	0.05	&	468	$\pm$	58			&	133	&	14	&	P14	&	549	$\pm$	77			&	4.29	$\pm$	0.06	&	a	&		498	$\pm$	46							&	\bf{	4.25	$\pm$	0.04				}	\\
30	&	Zn	&	4.56	$\pm$	0.05	&	G15	&	4.60	$\pm$	0.05	&	1122	$\pm$	140			&	309	&	4	&	P14	&	1241	$\pm$	50			&	4.65	$\pm$	0.02	&	a	&		1227	$\pm$	47							&	\bf{	4.64	$\pm$	0.02				}	\\
31	&	Ga	&	3.02	$\pm$	0.05	&	G15	&	3.06	$\pm$	0.05	&	32.4	$\pm$	4.0			&	9.62	&	6	&	P14	&	36.2	$\pm$	2.2			&	3.11	$\pm$	0.03	&	a	&		35.3	$\pm$	1.9							&	\bf{	3.10	$\pm$	0.02				}	\\
32	&	Ge	&	3.63	$\pm$	0.07	&	G15	&	3.67	$\pm$	0.07	&	132	$\pm$	23			&	32.6	&	9	&	P14	&	118	$\pm$	11			&	3.63	$\pm$	0.04	&	a	&		120	$\pm$	10							&	\bf{	3.63	$\pm$	0.03				}	\\
33	&	As	&	-			&	-	&	-			&	-					&	1.74	&	9	&	L09/P14	&	6.10	$\pm$	0.5			&	2.34	$\pm$	0.04	&	m	&		6.10	$\pm$	0.55							&	\bf{	2.34	$\pm$	0.04				}	\\
34	&	Se	&	-			&	-	&	-			&	-					&	20.3	&	7	&	L09/P14	&	67.5	$\pm$	4.7			&	3.38	$\pm$	0.03	&	m	&		67.5	$\pm$	4.7							&	\bf{	3.38	$\pm$	0.03				}	\\
35	&	Br	&	-			&	-	&	-			&	-					&	3.26	&	15	&	L09/P14	&	10.7	$\pm$	1.6			&	2.58	$\pm$	0.06	&	m	&		10.7	$\pm$	1.6							&	\bf{	2.58	$\pm$	0.06				}	\\
36	&	Kr	&	3.25	$\pm$	0.06	&	G15	&	3.29	$\pm$	0.06	&	55.0	$\pm$	8.3			&	5.22$\times10^{-5}$	&	-	&	L09/P14	&	1.64			$  \times$	$10^{-4}$	&	-2.23			&	p	&		55.0	$\pm$	8.3							&	\bf{	3.29	$\pm$	0.06				}	\\
37	&	Rb	&	2.47	$\pm$	0.07	&	G15	&	2.51	$\pm$	0.07	&	9.12	$\pm$	1.61			&	2.32	&	8	&	P14	&	7.12	$\pm$	0.57			&	2.41	$\pm$	0.03	&	a$^*$	&		7.35	$^{+	3.39	}_{-	0.79	}$				&	\bf{	2.42	$^{+	0.16	}_{-	0.05	}$	}	\\
38	&	Sr	&	2.83	$\pm$	0.06	&	G15	&	2.87	$\pm$	0.06	&	20.9	$\pm$	3.1			&	7.79	&	7	&	P14	&	23.3	$\pm$	1.6			&	2.92	$\pm$	0.03	&	a	&		22.8	$\pm$	1.4							&	\bf{	2.91	$\pm$	0.03				}	\\
39	&	Y	&	2.21	$\pm$	0.05	&	G15	&	2.25	$\pm$	0.05	&	5.01	$\pm$	0.62			&	1.46	&	5	&	P14	&	4.31	$\pm$	0.22			&	2.19	$\pm$	0.02	&	a	&		4.39	$\pm$	0.20							&	\bf{	2.19	$\pm$	0.02				}	\\
40	&	Zr	&	2.59	$\pm$	0.04	&	G15	&	2.63	$\pm$	0.04	&	12.0	$\pm$	1.2			&	3.63	&	5	&	P14	&	10.4	$\pm$	0.5			&	2.57	$\pm$	0.02	&	a	&		10.7	$\pm$	0.5							&	\bf{	2.58	$\pm$	0.02				}	\\
41	&	Nb	&	1.47	$\pm$	0.06	&	G15	&	1.51	$\pm$	0.06	&	0.912	$\pm$	0.137			&	0.283	&	10	&	P14	&	0.800	$\pm$	0.080			&	1.46	$\pm$	0.04	&	a	&		0.828	$\pm$	0.069							&	\bf{	1.47	$\pm$	0.03				}	\\
42	&	Mo	&	1.88	$\pm$	0.09	&	G15	&	1.92	$\pm$	0.09	&	2.34	$\pm$	0.54			&	0.961	&	10	&	P14	&	2.63	$\pm$	0.26			&	1.97	$\pm$	0.04	&	a	&		2.57	$\pm$	0.24							&	\bf{	1.96	$\pm$	0.04				}	\\
44	&	Ru	&	1.75	$\pm$	0.08	&	G15	&	1.79	$\pm$	0.08	&	1.74	$\pm$	0.35			&	0.69	&	5	&	P14	&	1.79	$\pm$	0.09			&	1.81	$\pm$	0.02	&	a	&		1.79	$\pm$	0.09							&	\bf{	1.80	$\pm$	0.02				}	\\
45	&	Rh	&	0.89	$\pm$	0.08	&	G15	&	0.93	$\pm$	0.08	&	0.240	$\pm$	0.049			&	0.132	&	5	&	P14	&	0.337	$\pm$	0.017			&	1.08	$\pm$	0.02	&	a$^*$	&		0.326	$^{+	0.027	}_{-	0.135	}$				&	\bf{	1.06	$^{+	0.03	}_{-	0.23	}$	}	\\[2pt]
46	&	Pd	&	1.55	$\pm$	0.06	&	G15	&	1.59	$\pm$	0.06	&	1.10	$\pm$	0.16			&	0.56	&	4	&	P14	&	1.38	$\pm$	0.06			&	1.69	$\pm$	0.02	&	a$^*$	&		1.352	$^{+	0.084	}_{-	0.421	}$				&	\bf{	1.68	$^{+	0.03	}_{-	0.16	}$	}	\\[2pt]
47	&	Ag	&	0.96	$\pm$	0.10	&	G15	&	1.00	$\pm$	0.10	&	0.282	$\pm$	0.073			&	0.201	&	9	&	P14	&	0.489	$\pm$	0.044			&	1.24	$\pm$	0.04	&	a$^*$	&		0.434	$^{+	0.099	}_{-	0.226	}$				&	\bf{	1.19	$^{+	0.09	}_{-	0.32	}$	}	\\
48	&	Cd	&	1.77	$\pm$	0.15	&	G15	&	1.81	$\pm$	0.15	&	1.82	$\pm$	0.75			&	0.674	&	7	&	L09/P14	&	1.57	$\pm$	0.11			&	1.75	$\pm$	0.03	&	a	&		1.58	$\pm$	0.11							&	\bf{	1.75	$\pm$	0.03				}	\\
49	&	In	&	0.80	$\pm$	0.20	&	G15	&	0.84	$\pm$	0.20	&	0.195	$\pm$	0.114			&	0.0778	&	5	&	P14	&	0.178	$\pm$	0.009			&	0.80	$\pm$	0.02	&	a	&		0.178	$\pm$	0.009							&	\bf{	0.80	$\pm$	0.02				}	\\
50	&	Sn	&	2.02	$\pm$	0.10	&	G15	&	2.06	$\pm$	0.10	&	3.24	$\pm$	0.84			&	1.63	&	15	&	L09/P14	&	3.60	$\pm$	0.54			&	2.11	$\pm$	0.06	&	a	&		3.50	$\pm$	0.46							&	\bf{	2.09	$\pm$	0.05				}	\\
51	&	Sb	&	-			&	-	&	-			&	-					&	0.145	&	14	&	P14	&	0.313	$\pm$	0.044			&	1.05	$\pm$	0.06	&	m	&		0.313	$\pm$	0.044							&	\bf{	1.04	$\pm$	0.06				}	\\
52	&	Te	&	-			&	-	&	-			&	-					&	2.28	&	7	&	L09/P14	&	4.69	$\pm$	0.33			&	2.23	$\pm$	0.03	&	m	&		4.69	$\pm$	0.33							&	\bf{	2.22	$\pm$	0.03				}	\\
53	&	I	&	-			&	-	&	-			&	-					&	0.53	&	20	&	L09/P14	&	1.10	$\pm$	0.22			&	1.59	$\pm$	0.08	&	m	&		1.10	$\pm$	0.22							&	\bf{	1.59	$\pm$	0.08				}	\\
54	&	Xe	&	2.24	$\pm$	0.06	&	G15	&	2.28	$\pm$	0.06	&	5.37	$\pm$	0.81			&	1.74$\times10^{-4}$	&	-	&	L09/P14	&	3.48			$  \times$	$10^{-4}$	&	-1.90			&	p	&		5.37	$\pm$	0.81							&	\bf{	2.28	$\pm$	0.06				}	\\
55	&	Cs	&	-			&	-	&	-			&	-					&	0.188	&	6	&	P14	&	0.371	$\pm$	0.022			&	1.12	$\pm$	0.03	&	m	&		0.371	$\pm$	0.022							&	\bf{	1.12	$\pm$	0.03				}	\\
56	&	Ba	&	2.25	$\pm$	0.07	&	G15	&	2.29	$\pm$	0.07	&	5.50	$\pm$	0.97			&	2.42	&	5	&	P14	&	4.63	$\pm$	0.23			&	2.22	$\pm$	0.02	&	a	&		4.67	$\pm$	0.22							&	\bf{	2.22	$\pm$	0.02				}	\\
57	&	La	&	1.11	$\pm$	0.04	&	G15	&	1.15	$\pm$	0.04	&	0.398	$\pm$	0.040			&	0.2414	&	3	&	P14	&	0.456	$\pm$	0.014			&	1.21	$\pm$	0.01	&	a$^*$	&		0.450	$^{+	0.020	}_{-	0.092	}$				&	\bf{	1.20	$^{+	0.02	}_{-	0.10	}$	}	\\
58	&	Ce	&	1.58	$\pm$	0.04	&	G15	&	1.62	$\pm$	0.04	&	1.17	$\pm$	0.12			&	0.6194	&	3	&	P14	&	1.16	$\pm$	0.03			&	1.62	$\pm$	0.01	&	a	&		1.16	$\pm$	0.03							&	\bf{	1.62	$\pm$	0.01				}	\\
59	&	Pr	&	0.72	$\pm$	0.04	&	G15	&	0.76	$\pm$	0.04	&	0.162	$\pm$	0.016			&	0.0939	&	3	&	P14	&	0.175	$\pm$	0.005			&	0.80	$\pm$	0.01	&	a	&		0.174	$\pm$	0.005							&	\bf{	0.79	$\pm$	0.01				}	\\
60	&	Nd	&	1.42	$\pm$	0.04	&	G15	&	1.46	$\pm$	0.04	&	0.813	$\pm$	0.081			&	0.4737	&	3	&	P14	&	0.862	$\pm$	0.026			&	1.49	$\pm$	0.01	&	a	&		0.857	$\pm$	0.025							&	\bf{	1.48	$\pm$	0.01				}	\\
62	&	Sm	&	0.95	$\pm$	0.04	&	G15	&	0.99	$\pm$	0.04	&	0.275	$\pm$	0.027			&	0.1536	&	3	&	P14	&	0.268	$\pm$	0.008			&	0.98	$\pm$	0.01	&	a	&		0.269	$\pm$	0.008							&	\bf{	0.98	$\pm$	0.01				}	\\
63	&	Eu	&	0.52	$\pm$	0.04	&	G15	&	0.56	$\pm$	0.04	&	0.102	$\pm$	0.010			&	0.05883	&	3	&	P14	&	0.102	$\pm$	0.003			&	0.56	$\pm$	0.01	&	a	&		0.102	$\pm$	0.003							&	\bf{	0.56	$\pm$	0.01				}	\\
64	&	Gd	&	1.08	$\pm$	0.04	&	G15	&	1.12	$\pm$	0.04	&	0.372	$\pm$	0.037			&	0.2069	&	3	&	P14	&	0.345	$\pm$	0.010			&	1.09	$\pm$	0.01	&	a	&		0.347	$\pm$	0.010							&	\bf{	1.09	$\pm$	0.01				}	\\
65	&	Tb	&	0.31	$\pm$	0.10	&	G15	&	0.35	$\pm$	0.10	&	0.0631	$\pm$	0.0164			&	0.03797	&	3	&	P14	&	0.0627	$\pm$	0.0019			&	0.35	$\pm$	0.01	&	a	&		0.0627	$\pm$	0.0019							&	\bf{	0.35	$\pm$	0.01				}	\\
66	&	Dy	&	1.10	$\pm$	0.04	&	G15	&	1.14	$\pm$	0.04	&	0.389	$\pm$	0.039			&	0.2558	&	3	&	P14	&	0.413	$\pm$	0.012			&	1.17	$\pm$	0.01	&	a	&		0.411	$\pm$	0.012							&	\bf{	1.16	$\pm$	0.01				}	\\
67	&	Ho	&	0.48	$\pm$	0.11	&	G15	&	0.52	$\pm$	0.11	&	0.0933	$\pm$	0.0270			&	0.05644	&	3	&	P14	&	0.0898	$\pm$	0.0027			&	0.51	$\pm$	0.01	&	a	&		0.0899	$\pm$	0.0027							&	\bf{	0.50	$\pm$	0.01				}	\\
68	&	Er	&	0.93	$\pm$	0.05	&	G15	&	0.97	$\pm$	0.05	&	0.263	$\pm$	0.033			&	0.1655	&	3	&	P14	&	0.260	$\pm$	0.008			&	0.97	$\pm$	0.01	&	a	&		0.260	$\pm$	0.008							&	\bf{	0.96	$\pm$	0.01				}	\\
69	&	Tm	&	0.11	$\pm$	0.04	&	G15	&	0.15	$\pm$	0.04	&	0.0398	$\pm$	0.0040			&	0.02609	&	3	&	P14	&	0.0405	$\pm$	0.0012			&	0.16	$\pm$	0.01	&	a	&		0.0405	$\pm$	0.0012							&	\bf{	0.16	$\pm$	0.01				}	\\
70	&	Yb	&	0.85	$\pm$	0.11	&	G15	&	0.89	$\pm$	0.11	&	0.219	$\pm$	0.063			&	0.1687	&	3	&	P14	&	0.256	$\pm$	0.008			&	0.96	$\pm$	0.01	&	a	&		0.255	$\pm$	0.008							&	\bf{	0.96	$\pm$	0.01				}	\\
71	&	Lu	&	0.10	$\pm$	0.09	&	G15	&	0.14	$\pm$	0.09	&	0.0389	$\pm$	0.0090			&	0.02503	&	3	&	P14	&	0.0375	$\pm$	0.0011			&	0.13	$\pm$	0.01	&	a	&		0.0376	$\pm$	0.0011							&	\bf{	0.12	$\pm$	0.01				}	\\
72	&	Hf	&	0.85	$\pm$	0.05	&	G15	&	0.89	$\pm$	0.05	&	0.219	$\pm$	0.027			&	0.1065	&	3	&	P14	&	0.157	$\pm$	0.005			&	0.75	$\pm$	0.01	&	a$^*$	&		0.158	$^{+	0.088	}_{-	0.006	}$				&	\bf{	0.75	$^{+	0.19	}_{-	0.02	}$	}	\\
73	&	Ta	&	-			&	-	&	-			&	-					&	0.015	&	10	&	P14	&	0.0218	$\pm$	0.0022			&	-0.11	$\pm$	0.04	&	m	&		0.0218	$\pm$	0.0022							&	\bf{	-0.11	$\pm$	0.04				}	\\
74	&	W	&	0.83	$\pm$	0.11	&	G15	&	0.87	$\pm$	0.11	&	0.209	$\pm$	0.061			&	0.096	&	10	&	P14	&	0.137	$\pm$	0.014			&	0.69	$\pm$	0.04	&	a$^*$	&		0.141	$^{+	0.129	}_{-	0.017	}$				&	\bf{	0.70	$^{+	0.28	}_{-	0.06	}$	}	\\
75	&	Re	&	-			&	-	&	-			&	-					&	0.04	&	5	&	P14	&	0.056	$\pm$	0.003			&	0.31	$\pm$	0.02	&	m	&		0.0564	$\pm$	0.0028							&	\bf{	0.30	$\pm$	0.02				}	\\
76	&	Os	&	1.40	$\pm$	0.05	&	G15	&	1.44	$\pm$	0.05	&	0.776	$\pm$	0.097			&	0.495	&	5	&	P14	&	0.683	$\pm$	0.034			&	1.39	$\pm$	0.02	&	a	&		0.693	$\pm$	0.032							&	\bf{	1.39	$\pm$	0.02				}	\\
77	&	Ir	&	1.42	$\pm$	0.07	&	G15	&	1.46	$\pm$	0.07	&	0.813	$\pm$	0.144			&	0.469	&	5	&	L09/P14	&	0.640	$\pm$	0.032			&	1.36	$\pm$	0.02	&	a$^*$	&		0.649	$^{+	0.308	}_{-	0.040	}$				&	\bf{	1.36	$^{+	0.17	}_{-	0.03	}$	}	\\
78	&	Pt	&	-			&	-	&	-			&	-					&	0.925	&	5	&	P14	&	1.24	$\pm$	0.06			&	1.65	$\pm$	0.02	&	m	&		1.24	$\pm$	0.06							&	\bf{	1.65	$\pm$	0.02				}	\\
79	&	Au	&	0.91	$\pm$	0.08	&	G15	&	0.95	$\pm$	0.08	&	0.251	$\pm$	0.051			&	0.148	&	12	&	P14	&	0.197	$\pm$	0.024			&	0.85	$\pm$	0.05	&	a	&		0.207	$\pm$	0.021							&	\bf{	0.87	$\pm$	0.04				}	\\
80	&	  Hg	&	-			&	-	&	-			&	-					&	0.35	&	50	&	P14	&	0.458	$\pm$	0.229			&	1.21	$\pm$	0.18	&	m	&		0.458	$\pm$	0.229$^l$							&	\bf{1.21 $\pm$ 0.18}$^l$	\\
81	&	 Tl	&	0.90	$\pm$	0.20	&	G15	&	0.94	$\pm$	0.20	&	0.245	$\pm$	0.144			&	0.14	&	11	&	P14	&	0.180	$\pm$	0.020			&	0.81	$\pm$	0.05	&	a	&		0.181	$\pm$	0.020							&	\bf{	0.81	$\pm$	0.04				}	\\
82	&	  Pb	&	1.92	$\pm$	0.08	&	G15	&	1.96	$\pm$	0.08	&	2.57	$\pm$	0.52			&	2.62	&	8	&	P14	&	3.32	$\pm$	0.266			&	2.07	$\pm$	0.03	&	a	&		3.17	$\pm$	0.24							&	\bf{	2.05	$\pm$	0.03				}	\\
83	&	  Bi	&	-			&	-	&	-			&	-					&	0.11	&	9	&	L09/P14	&	0.138	$\pm$	0.012			&	0.69	$\pm$	0.04	&	m	&		0.138	$\pm$	0.012							&	\bf{	0.69	$\pm$	0.04				}	\\
90	&	  Th	&	0.03	$\pm$	0.10	&	G15	&	0.07	$\pm$	0.10	&	0.0331	$\pm$	0.0086			&	0.03	&	7	&	P14	&	0.0339	$\pm$	0.0024			&	0.08	$\pm$	0.03	&	a	&		0.0339	$\pm$	0.0023							&	\bf{	0.08	$\pm$	0.03				}	\\
92	&	  U	&	-			&	-	&	-			&	-					&	0.0081	&	7	&	P14	&	0.0089	$\pm$	0.0006			&	-0.50	$\pm$	0.03	&	m	&		0.0089	$\pm$	0.0006							&	\bf{	-0.50	$\pm$	0.03				}	\\
			
		\end{longtable}
	}
\end{landscape}

\section{Devolatilization and the Volatility Trend of Bulk Earth}
\label{sec:pattern}

\subsection{Compositional comparison between the bulk Earth and the proto-Sun}
\label{sec:pattern1}
\begin{figure}[!ht]
	\begin{center}
		\includegraphics[trim=2.5cm 1.2cm 2.5cm 1.5cm, scale=0.56,angle=90]{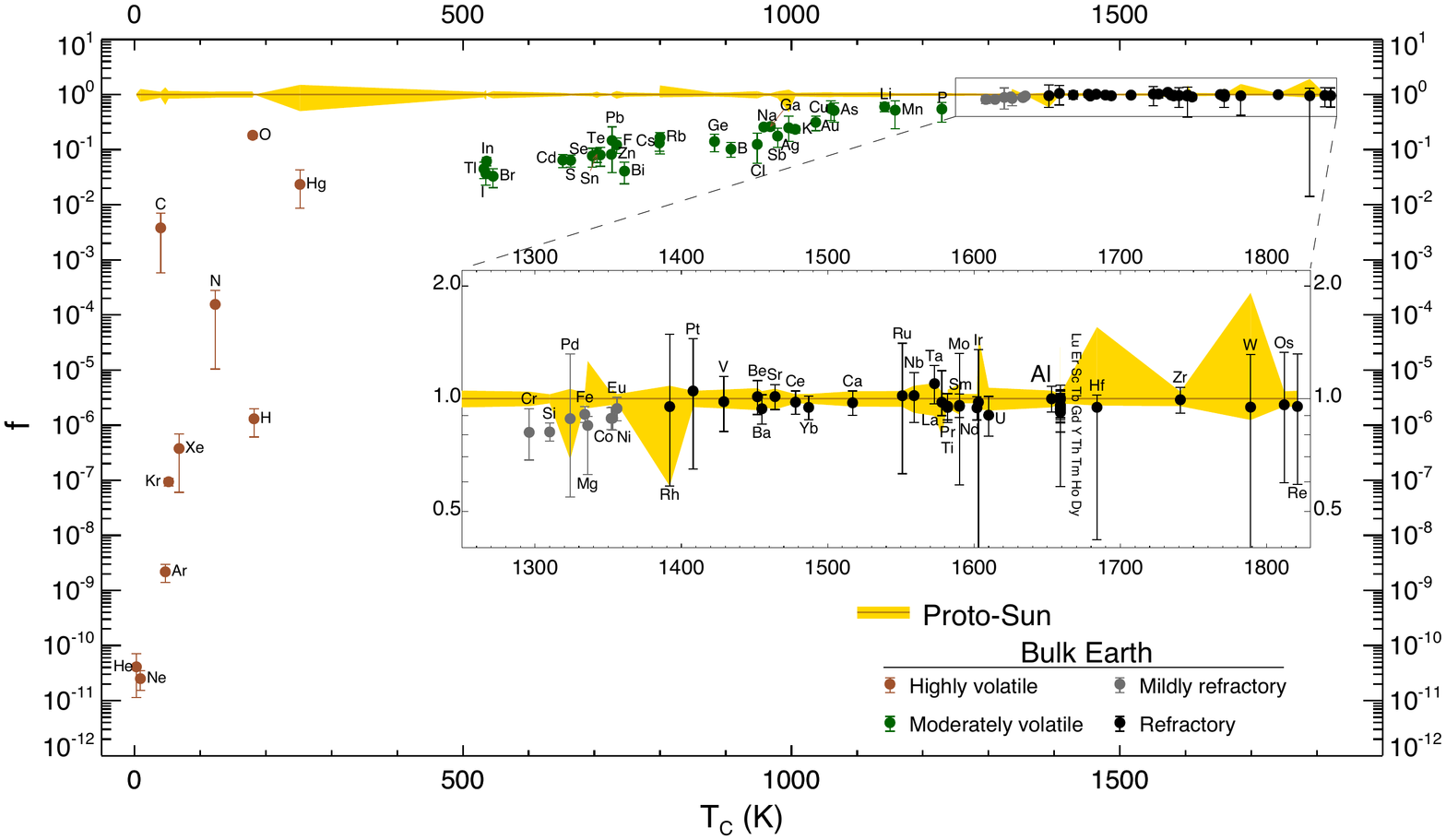} 
		
		\caption{The abundance ratios $f$, of the bulk Earth to proto-Sun where $f = \mathrm{N(X)_{Earth} / N(X)_{Sun}}$, normalized to Al.
			$f$ is plotted in order of increasing 50\% condensation temperature $T_C$ \citep{Lodders2003}.
			Compared with the Sun, the Earth is depleted in the most volatile elements. The abundances of the refractory elements in the Sun and Earth are 
			indistinguishable. Elements are classified into four categories: \textit{highly volatile} ($T_C < 500 $ K), 
			\textit{moderately volatile}  (500 K $ < T_C < $ 1250 K), \textit{mildly refractory} (1250 K $< T_C < $1390 K) and 
			\textit{refractory} ($T_C > 1390 $ K).}		
		\label{fig:ErtSunAbu}
	\end{center}
\end{figure}

Our goal is to quantify as precisely and accurately as possible the chemical relationship between Earth and Sun.
For Sun, we use the protosolar abundances described in the previous section.
For Earth we use our recent bulk Earth elemental abundances \mbox{\citep{Wang2018}}.
To compare these two data sets, we normalize the protosolar abundances and the bulk Earth abundances to Al (i.e., $N(\mathrm{Al})=10^6$).
A normalization to Al (rather than Mg or Si) is important because the condensation temperature of Al is the highest of any of the major elements in a gas of solar composition, and 
our main goal is to assess the depletion in elemental abundances as a function of condensation temperature.
The abundance ratios ($f$) of the bulk Earth to the proto-Sun (i.e., $f = N\mathrm{(X)_{Earth}} / N\mathrm{(X)_{Sun}}$) 
is plotted in order of increasing 50\% condensation temperature ($T_C$) in Fig. {\ref{fig:ErtSunAbu}}. 
The $T_C$ estimates are from \cite{Lodders2003}. 
Fig. \ref{fig:ErtSunAbu} shows that bulk Earth abundances are systematically lower than or equal to protosolar abundances and that there is a definite pattern in the abundance ratios as a function of $T_C$.
  
Elements are split into four discrete categories in Fig. {\ref{fig:ErtSunAbu}}.
At high condensation temperatures ($T_C >$ 1390  K), the abundance ratios are unity $(f \approx 1$), indicating that refractory elements have not been depleted. 
The abundance ratios begin to decrease for mildly refractory elements (1250 K $< T_C < $ 1390 K) (see the 8 grey points on the left side of the inset in Fig. {\ref{fig:ErtSunAbu}}).
The abundance ratios of moderately volatile elements (500 K $ < T_C < $ 1250 K) progressively decrease in a fairly tight linear correlation with decreasing $T_C$.
Below 500 K, the tight linear correlation weakens, but the positive correlation between decreasing condensation temperatures and decreasing abundance ratios is still strong. 

\subsection{Quantification of the \textquotedblleft bulk" volatility trend (VT)}
\label{sec:pattern2}

\begin{figure}[!ht]	
	\begin{center}
		\includegraphics[trim=0.5cm 2.5cm 0.5cm 3.0cm, scale=0.67,angle=90]{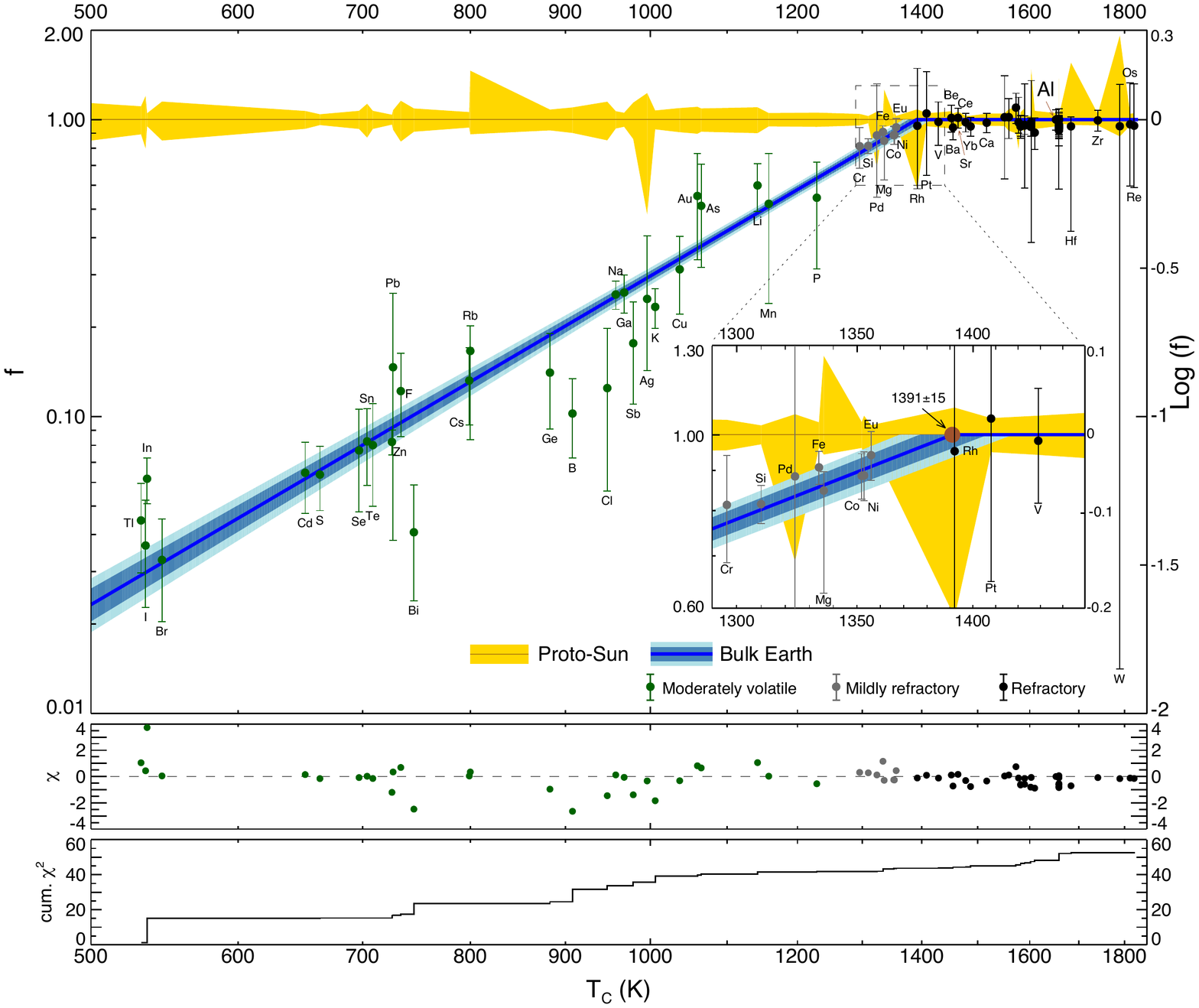} 
		\caption{Comparison of the elemental abundances of the Earth and Sun. This plot is similar to Fig. \ref{fig:ErtSunAbu} except here, 
			we only plot elements with $T_C > 500$ K and the $T_C$ axis is logarithmic. 
			The diagonal dark blue line and its horizontal continuation in the upper right, is our best $\chi^2$ fit to the data.
			The dark and light blue diagonal regions are the 68\% and 95\% confidence intervals on the fit.
			For details see \mbox{\ref{sec:app4}}. 	
			The diagonal dark blue line reaches $f=1$ at the critical devolatilization temperature $T_{\mathrm{D}}(\mathrm{E}) =1391 \pm 15$ K.
			The inset shows more of the details of this region. 
			The lower two panels show the values of $\chi$ and the cumulative $\chi^2$ as a function of $T_C$.}			
		\label{fig:TEDP} 		
	\end{center}
\end{figure}

The Earth-to-Sun abundance ratios ($f$) are plotted in Fig. \ref{fig:TEDP} as a function of $T_C$ (for $T_C > 500$ K).
An important result of quantifying the volatility trend 
is estimating the temperature that separates the region of undepleted terrestrial abundances on the right (i.e. $f \approx 1$) from the region on the left where $f$ decreases with decreasing
condensation temperature. 
The critical devolatilization temperature for Earth $T_{D}(\mathrm{E})$ is then derived from a simultaneous model fit to depleted abundances and undepleted abundnaces.

In brief, we perform a $\chi^2$ fit of the $f$ values to the joint model:
$\log(f) = \alpha\log(T_C) + \beta$ and $\log(f) = 0$.   
The best-fit coefficients that we obtain are $\alpha=3.676 \pm 0.142$ and $\beta=-11.556 \pm 0.436$. The reduced $\chi^2$ of the best fit is about 1.2. 
Since $T_{\mathrm{D}}(\mathrm{E}) = 10^{-\beta/\alpha}$, these coefficients yield the best-fit devolatilization temperature $T_{\mathrm{D}}(\mathrm{E}) = 1391 \pm 15$ K. Details of the minimization process (and goodness-of-fit test) are given in \mbox{\ref{sec:app4}}. 
The linear relationship shown in the log-log plot of Fig. \ref{fig:TEDP} may not be the only possible relationship, but there are very few relationships simpler than linear. It is the simplest relationship we could find, and has an acceptable goodness-of-fit (see lower two panels of Fig. \ref{fig:TEDP} as well as \ref{sec:app4}).

The quantification of the VT provides an empirical observation upon which interpretation of the devolatilization processes can be based. The systematic depletion of moderately volatile elements versus their condensation temperatures demonstrates that the devolatilization processes are largely volatility controlled, which is consistent with a comparison of various meteorites to CI chondrites \citep{Bland2005}. 

The critical devolatilization temperature $T_{\mathrm{D}}(\mathrm{E})$ is plausibly interpreted as the highest temperature experienced by the material in the feeding zone of the proto-Earth. 
More specifically, magnesium silicates are the dominant condensed phase(s) and have an effective 
condensation temperature of $\sim T_{\mathrm{D}}(\mathrm{E})$.
During repeated transient heating events, the latent heat of vaporization of the abundant mineral oxides could prevent the high-temperature vaporization and depletion of more refractory material. Hence, producing the $f \sim 1$ region of the VT.
Less refractory elements with $T_C < 1391 K$ would not be protected by the latent heat of the dominant phase and would
be susceptible to vaporization and depletion, depending on their $T_C$. 
We have not propagated the uncertainties on the $T_C$ estimates of \cite{Lodders2003} (see discussion in Section \ref{sec:discu1}).

The canonical explanation for this depletion of volatiles is the partial condensation of hot gas being cooled \citep{Wasson1974, Cassen1996, Palme2003, Davis2006, Davis2014}. In this scenario, a large hot ($>$ 1800 K) reservoir of solar composition gas needs to cool down to $\sim$ 1390 K and then, what has condensed needs to be partially separated from the original gas that has not condensed. However, the only location in recent protoplanetary disk models that is hot enough to allow partial condensation from temperatures $>$ 1800 K, is in the inner disk, very close to the Sun. The expulsion and cooling of this material can explain some of the processing necessary to account for the mixture of refractory and volatile material seen in some meteoritic material, but it seems implausible as a unified mechanism to explain the linear relation in Fig. \ref{fig:TEDP} because of the unlikelihood that a significant fraction of the mass of the Earth could be so processed and then transported. 

Our preferred devolatilization process involves the partial sublimation of cold, volatile-rich material being heated (some of it up to $\sim$ 1390 K) as it accretes through the midplane of the protoplanetary disk and ends up in the eventual feeding zone of the Earth.
In the partial condensation model, all the material needs to be hotter than $\sim$ 1800 K as an initial condition (also see \cite{Albarede2009} for an explanation of the kink of the volatility trend corresponding to the remarkable accretion hiatus at the end of major element condensation). In the partial sublimation model, the dominant material (e.g. silicate grains) needs to reach an average maximum temperature of $\sim$ 1390 K. This lower (average) maximum temperature seems like a more plausible scenario that is more consistent with current protoplanetary disk models \citep[e.g.][]{Dullemond2010, Morbidelli2015b, Carrera2017}. However, these two models are not mutually exclusive. Both devolatilization processes could have contributed to the depletion. 

\section{Discussion}
\label{sec:discu}

\subsection{The volatility trend of Earth materials}
\label{sec:discu1}

A volatility trend (VT) in the composition of Earth materials has been noted previously in the literature, e.g. \cite{Kargel1993}, \cite{McDonough2014}, and \cite{Palme2014b}.
Our assessment of the VT differs from previous research in a number of ways. Our modeling is based on normalization to a revised set of photospheric abundances while previous estimates of the VT are based primarily on bulk analysis of chondrites. We have also modeled the bulk composition of the Earth \citep{Wang2018}, as opposed to earlier work based on the Bulk Silicate Earth (BSE).
BSE modeling is useful for estimating the depletion of siderophile elements from the mantle into the core, but is less useful for estimating the depletion of elements in the bulk Earth compared to their protosolar abundances.
Previous work has semi-qualitatively estimated the VT based on visual fitting of the abundance - temperature correlation. A key point of our work is that it represents fitting of data with consideration of uncertainties. Hence, we can assess statistically the nature of deviations from the line. 

Our model shows that the abundances of elements with condensation temperatures between ca. 500 and \mbox{1400 K} can be satisfactorily fit by a linear correlation between log abundance and log temperature. The correlation is generally interpreted as relating to gas-solid fractionation of the elements such as would occur during evaporation or condensation. 

The condensation temperatures used in this work are derived from \mbox{\cite{Lodders2003}}. These 50\% condensation temperatures are estimated from where 50\% of the element passes from gaseous to solid phase, either directly as elements, compounds, or as substitutions into condensing phases at a total pressure of $10^{-4}$ bar. Therefore, our quantification of the VT does not take into account the pressure and model dependencies of elemental condensation temperatures. The pressure dependencies of condensation temperatures \citep{Ebel2000, Ebel2006} are highly correlated in that they decrease and increase in unison. Therefore,  pressure dependence will not have a significant effect on the linearity of the blue curve in Fig. 5. 
The model dependencies of $T_C$, in particular of trace elements, are discussed in \cite{Lodders2003} (also see \cite{Wai1977} cited therein). 

Mg or Si have been the elements of choice for normalizing the elemental abundances of the Earth and CI chondrites. This is a natural consequence of the abundances of these elements and their wide use in understanding geochemical fractionation in magmatic systems. For example, both \cite{Palme2014b} and \cite{McDonough2014} chose Mg as the normalizing reference element. However, in terms of cosmochemistry Mg and Si are only mildly refractory and so can experience both magmatic fractionation, e.g. crystallization of forsterite (Mg$_2$SiO$_4$) in chondrules, as well as cosmochemical fractionation due to volatility in high temperature processes. Similarly, the variability of Mg/Si ratios in meteoritic material (with respect to the bulk Earth) may also be due to the loss or addition of early condensed forsterite, as well as to the small (and not-necessarily-representative) sample sizes of meteoritic material. The respective $T_C$ values of Mg and Si are 1330 K and 1310 K, both of which are less than $T_{\mathrm{D}}(\mathrm{E})=1391\pm15$ K.
Thus, both Mg and Si are slightly depleted in the Earth relative to CI chondrites. 
Thus, normalizing the Earth and CI chondrites  to Mg or Si, produces an apparent enrichment of the Earth in refractory lithophiles by a factor of about $\sim 1.2$ (normalized to Mg) or $\sim 1.4$ (normalized to Si) (see, however, \cite{Hezel2008a} and \cite{Hezel2008b} for an alternative view based on the enrichment of CAIs in other types of carbonaceous chondrites relative to CI chondrites). In this work we have chosen to use Al as the normalizing element because it is the most refractory of the major rock-forming elements and hence is less susceptible to fractionation from volatility.

\cite{Kargel1993} compared the Earth's abundances (in ppm) to CI chondritic abundances (in ppm).
However, since the Earth is much more depleted in volatiles than are CI chondrites, this procedure elevates the abundances of refractories compared to
the total mass by a factor of $\sim 2.8$. This apparent enrichment is due to the depletion of volatiles from the total mass of the Earth.

\begin{figure}[!ht]	
	
	\begin{center}
		\includegraphics[trim=1.0cm 0.8cm 1.2cm 1.4cm, scale=0.6,angle=90]{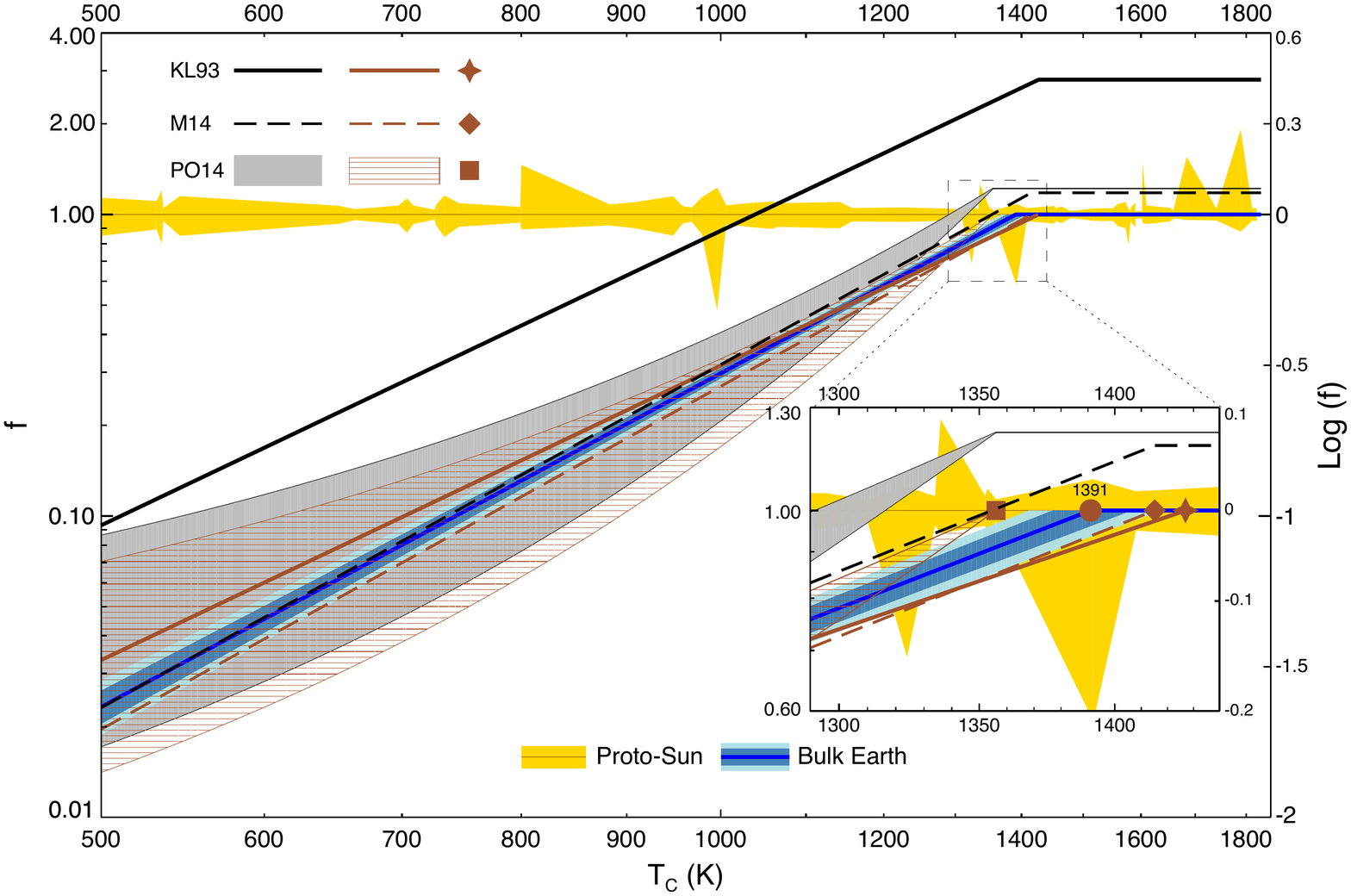} 
		\caption{Comparison of our best-fit volatility trend (VT, 
			in blue from Fig. \ref{fig:TEDP}) with the previous analogous VTs (i.e. volatility trends) of \cite{Kargel1993} (KL93), \cite{McDonough2014} (M14) and \cite{Palme2014b} (PO14).  These previous DPs are based on the normalization of 
			bulk silicate Earth to CI chondrites (as a proxy for the Sun).
			The curved grey wedge represents the range from Fig. 21 of \mbox{\cite{Palme2014b}}.
			To make these previous VTs more comparable to our analysis, we renormalize them to 1 on their horizontal right hand sides.
			The renormalized curves are shown as brown versions of the original non-renormalized line style. 
			For details see footnotes of Table \mbox{\ref{tab:patternfunc}}.	
			This procedure yields $T_{\mathrm{D}}(\mathrm{E})$ values for each curve (see inset).
			These values can be found in Table \mbox{\ref{tab:patternfunc}}.
		}
		\label{fig:TEDPcompare} 		
	\end{center}
\end{figure}

To plot these previous VTs and make them more comparable with our work: (1) we estimate the coefficients $\alpha$ and $\beta$ of VTs based on published figures in
\cite{Kargel1993} (KL93), \cite{McDonough2014} (M14) and \cite{Palme2014b} (PO14); (2) we renormalize these VTs to remove apparent lithophile enrichment (see details in Table \mbox{\ref{tab:patternfunc}}).
The comparison between our VT and these previous analogous VTs before and after renormalization is illustrated in Fig. \ref{fig:TEDPcompare}.
The analogous VT of KL93 (after normalization) is essentially out of the 2$\sigma$ range of our VT, with a less steep slope and a significantly higher $T_D$ = 1427 K. The analogous VT of M14 (after normalization) basically aligns the lower envelope of the 2$\sigma$ range of our VT, with a comparable slope and a higher $T_D$ = 1415 K. The analogous VT of PO14 (after normalization) largely overlaps with our VT, but appears to be less constrained, with a significantly lower $T_D$ = 1356 K.


\subsection{The devolatilization of the inner solar system}
\label{sec:discu1-2}
The VT indicates a simple relationship between element condensation temperature and element abundance, in a specific range of condensation temperatures.  Using the best available data, the elemental abundances in the designated condensation temperature range are well fit by a linear relationship. Such a correlation is quite remarkable.  It could represent a large number of sweeps through the temperature range and the VT reflects the average of very many evaporation and condensation cycles.

The evident devolatilization has likely affected the planetesimal protoliths accreting to form planet Earth.  As such the process(es) responsible could have affected all scales of objects from chondrules  (sub mm) through to chondrite parent bodies and larger differentiated asteroids.  Hence during pebble accretion and at each stage of oligarchic growth to planetesimals and planets, mixtures of materials potentially with different thermal histories could be present.

Chondrites as a whole are taken as the building blocks of the solar system and there is remarkably little difference among the compositions of ordinary chondrites. 
This agreement suggests that not only was the solar nebula well mixed, but that chondrule formation itself did not result in large scale fractionation and separation of moderately volatile elements.

The other high-temperature component in chondrites, and especially carbonaceous chondrites, are refractory Ca-Al-rich inclusions (CAIs).  Many CAIs have experienced melting at similar temperatures to chondrules, at ca. 1800 K.  But unlike chondrules, CAIs typically show fractionations of even refractory elements.  A feature of many CAIs is the variability in the abundances of the more volatile rare-earth elements (REEs), e.g. Ce, Eu and Yb.  These are typical depletions, such as in the Group III inclusions of \cite{Mason1982}, 
but enrichments are also found for example in hibonite-bearing inclusions \citep{Ireland1988, Ireland2000}. 
Such depletions can be interpreted as being due to evaporation of these elements, or as partial condensation where the CAI precursors are isolated from the condensing gas prior to full condensation of lithophile elements from the gas.  On the other hand, the Group II inclusions of \cite{Mason1982} 
show depletion of the most refractory REEs and other trace elements; these can only be interpreted as a result of partial condensation.  Hence evaporation and condensation of the most refractory elements are likely taking place in the inner solar system.  

This has profound implications for the structure and thermal history of the solar nebula.  As evidenced by the CAIs, dust carrying refractory elements into the inner solar system has completely evaporated to the gas phase, yet is retained in the nebula rather than accreting to the Sun - cf. \cite{Gonzalez2014}. We are familiar with the concept of the snow line indicating the gas-solid transition for water, but there must also be an analogous evaporation line, closer to the Sun, that indicates the front of complete evaporation of refractory lithophile elements.  Further out from this, refractory lithophile elements are not (completely) evaporated with devolatilization dependent on the condensation temperature of the individual elements.

\subsection{Extrapolating the {quantified VT} to lower condensation temperatures}
\label{sec:discu2}
\begin{figure} [!ht]	
	\begin{center}
		\includegraphics[trim=6cm 2.0cm 0cm 3cm, scale=0.6,angle=90]{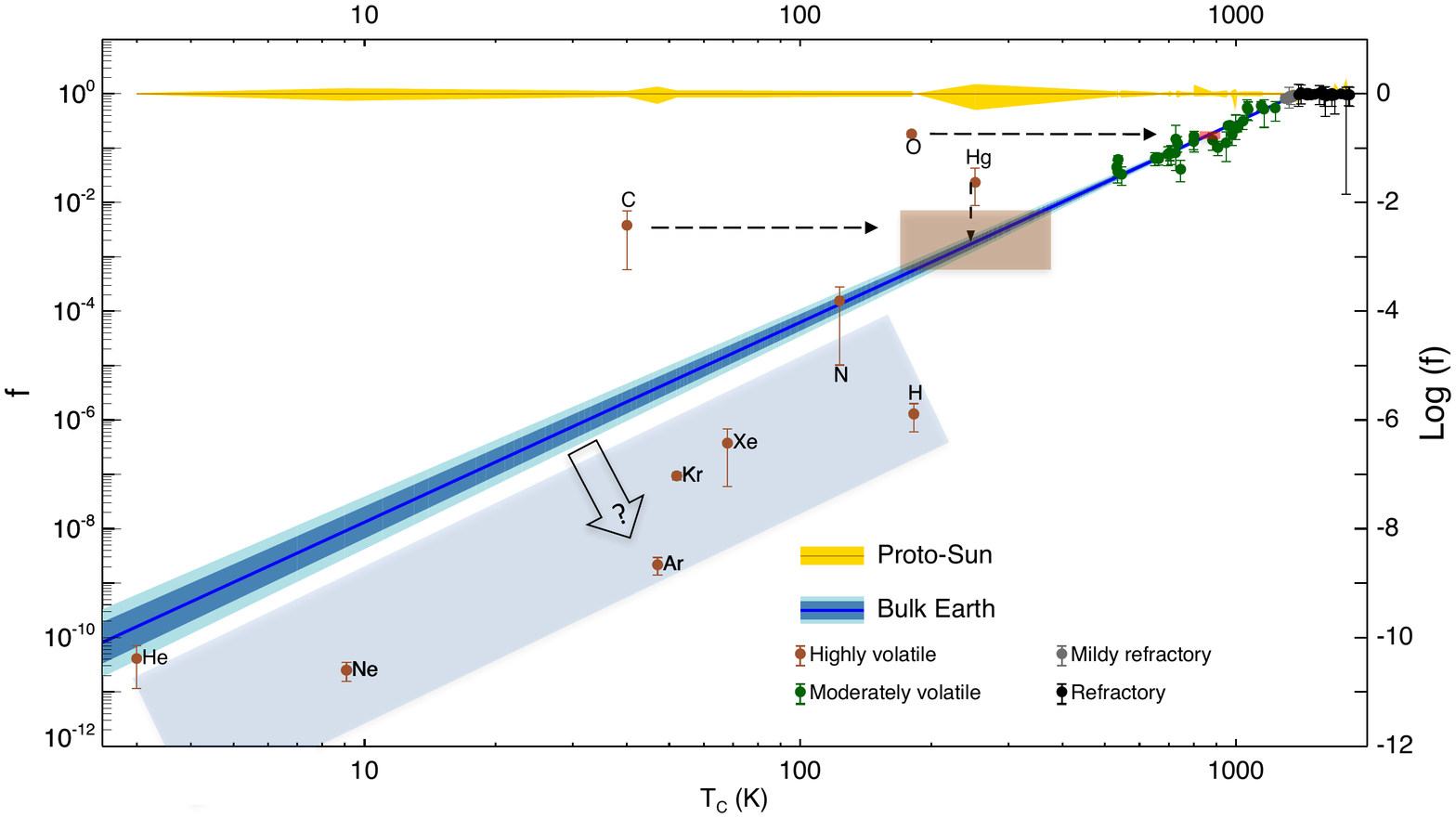} 
		\caption{Extrapolation of our VT to the realm of highly volatile elements ($T_C < 500 K$). 
			The position of the light blue rectangle indicates the stronger depletion of noble gases and hydrogen. 
			For C and O, the horizontal arrows pointing to the brown and red boxes, indicate the modification in $T_C$ values that would bring their 
			abundances into agreement with our best-fit VT. 
			The vertical dashed arrow beneath Hg indicates the change in the Earth's relative abundance if the Sun's abundance of Hg were raised by
			a factor of $13 \pm 7$.
			For discussion see Section \ref{sec:discu2}.}
		\label{fig:TEDPentire} 		
	\end{center}
\end{figure} 
The VT has been constrained by fits to elements with $T_C > 500$ K but the linear trend can be extrapolated to lower condensation temperatures (see Fig. \mbox{\ref{fig:TEDPentire}}). Nitrogen lies on the extrapolation of the VT but the other highly volatile elements scatter around it. Hydrogen and the noble gases are more depleted relative to the model, 
while carbon and oxygen appear more enriched. Interestingly, mercury (Hg) is also significantly above the extrapolation of the VT. The implications for the behavior of these elements are discussed below.

\subsubsection{The Extra Depletion of Noble Gases and Hydrogen}  
\label{sec:discu2a} 

Compared to the extrapolation of our VT below $T_C = 500$ K, there seems to be an extra depletion for elements with $T_C $ less than that of water ($T_C = 182$ K) (see Fig. \ref{fig:TEDPentire}). 
One plausible explanation for this extra depletion could be due to the latent heat of water ice which is the dominant phase at low temperatures.
During repeated transient heating events, the latent heat of water ice would buffer heat excursions above $ T \sim 182$ K but not at lower temperatures.
Thus, highly volatile elements with $T_C < 182$ K would be repeatedly susceptible to extra sublimation during these low temperature heating events.

It is also plausible that the transport of noble gases and hydrogen into the proto-Sun led to the primordial depletion of these gases from the dust that coalesced to form planetesimals. 
H and noble gases are unreactive with cold dust, and once separated from it, their history is not associated with dust. 
Thus, hydrogen and the noble gases are not entrained in the solid precursors of the terrestrial planets and are subject to an extra depletion than one based solely on their condensation temperatures.

\subsubsection{C and O} 
\label{sec:discu2b}
Under the assumption of equilibrium condensation from an initially hot, solar-composition gas, \cite{Lodders2003} 
reported the 50\% condensation temperatures of carbon ($T_C(\mathrm{C}) = T_C(\mathrm{CH_4}) \sim 40$ K) and oxygen ($T_C(\mathrm{O}) = T_C(\mathrm{H_2O}) \sim 180$ K).
With these $T_C$ values, the abundances of carbon and oxygen appear 2-3 orders of magnitude higher than the extrapolation of our VT below $T_C = 500$ K, would
predict (Fig. \ref{fig:TEDPentire}).

We attribute these higher abundances to the significant fractions of carbon and oxygen that are in refractory phases: kerogen and graphite for carbon;
silicates and mineral oxides for oxygen. To estimate these fractions, we make the approximation that both carbon and oxygen are distributed between volatile ($f_\mathrm{vol}$) and refractory ($f_\mathrm{ref}$) components satisfying $f_\mathrm{vol} + f_\mathrm{ref} = 1$. We adopt approximate condensation temperatures for each of these components (see Table \ref{tab:TcCO}). 
With these two approximations, we can use the volatility trend in Fig \ref{fig:TEDPentire}, to estimate the fractions of these two components that are missing from the nebular composition: $f_\mathrm{vol,missing}$ and $f_\mathrm{ref,missing}$. For both carbon and oxygen we use the approximation that $f_\mathrm{vol,missing} = 1.00$, since in the $T_C$ range below 200 K the Earth-to-Sun abundance ratios are less than $10^{-3}$.
The abundances of carbon and oxygen in the Earth tell us the total fraction of each element that has gone missing ($f_\mathrm{total,missing}$). 
Thus, we have the equation
$f_\mathrm{total,missing} = f_\mathrm{vol} f_\mathrm{vol, missing} + f_\mathrm{ref} f_\mathrm{ref,missing}$.
For both carbon and oxygen, we can now solve these two equations for the two unknowns, ($f_\mathrm{vol},f_\mathrm{ref}$). 
For C-bearing volatile and refractory phases we find ($0.91\pm 0.08$, $0.09 \pm 0.08$).
For O-bearing volatile and refractory phases we find ($0.80\pm 0.04$, $0.20 \pm 0.04$).
See Table \mbox{\ref{tab:TcCO}} for details.

Both C and O have significant abundances in refractory phases. Therefore, we expect both to have significantly higher abundances than would be the case if 100\% of the element were in the volatile phase. The blue line in Fig. \ref{fig:TEDPentire} (at the $T_C$ values of the volatile phases of C and O), is a prediction of their abundance \textit{if} they were 100\% in the volatile phase. We obtain an effective condensation temperature of C and O by horizontally shifting their abundances to agree with the best fit VT. 
The corresponding effective $T_C$ for C and O are illustrated by the brown and red boxes respectively shown in Fig. \ref{fig:TEDPentire}. 
We find $T_{C,eff}$(C) = $305^{+73}_{-135}$ K and  $T_{C,eff}$(O) =  $875 \pm 45$ K. 
These corrections enable this Sun/Earth VT to be potentially useful for estimating carbon and oxygen abundances in other rocky bodies from their host's stellar abundances.

\begin{table*}[!htbp]
	{\small
		\caption{Fractional distribution of carbon and oxygen between volatile and refractory phases ($f_\mathrm{vol}$, $f_\mathrm{ref}$) in the solar nebular material.}
		\begin{center}
			\begin{tabular}{cccc}
				\toprule
				&& Carbon & Oxygen\\
				\hline
				$f_\mathrm{total, missing}$$^{a}$&& $0.996\pm0.003$ & $0.818\pm0.023$ \\
				\hline
				\textbf{Volatile phases}$^b$  &&  CH$_4$, CO$_2$, CO  & H$_2$O, CO$_2$, CO \\ 
				Adopted $T_C$ (K) && $47\pm22$ & $103\pm79$ \\
				$f_\mathrm{vol,missing}$$^c$&& 1.00 & 1.00 \\
				\bm{$f_\mathrm{vol}$} && \bm{$0.91\pm0.08$} & \bm{$0.80\pm0.04$}\\
				
				\hline  
				\textbf{Refractory phases}$^b$  &&  Kerogen, graphite  & Silicates/mineral oxides \\
				Adopted $T_C$ (K) && $600\pm30$ & $1350\pm40$ \\
				$f_\mathrm{ref,missing}$$^c$&&  $0.955\pm0.014$ & $0.106^{+0.120}_{-0.106}$ \\
				\bm{$f_\mathrm{ref}$} $(=1-f_\mathrm{vol})$ && \bm{$0.09\pm0.08$} & \bm{$0.20\pm0.04$}\\													
				\bottomrule
				\multicolumn{4}{p{10cm}}{\scriptsize $^a$ The total fraction of each element that has gone missing from the solar nebula, calculated by 1 - $f$ of C and O as shown in Fig. \ref{fig:TEDPentire} (where f is the Earth-to-Sun abundance ratio).}\\
				\multicolumn{4}{p{10cm}}{\scriptsize $^b$ The respective condensation temperatures ($T_C$) of these C- or O-bearing species refer to \cite{Lodders2003} for CH$_4$ (41 K), H$_2$O (182 K), graphite (626 K), and major silicates/mineral oxides including enstatite (1316 K), forsterite (1354 K), diopside (1347 K), anorthite (1387 K), and spinel (1397 K); \cite{Lewis1980} for CO (25 K) and CO$_2$ (69 K); \cite{Pizzarello2010} for kerogen (573 K).}\\
				\multicolumn{4}{p{10cm}}{\scriptsize $^c$ The fractions of the volatile and refractory components that have gone missing from the solar nebula, inferred from the volatility trend in Fig. \ref{fig:TEDPentire}, using the adopted $T_C$ values.}\\						
			\end{tabular}
		\end{center}
		\label{tab:TcCO}
	}
\end{table*}

\subsubsection{Hg}
\label{sec:discu2c}

There are no observable Hg photospheric lines.
Mercury has a condensation temperature of 252 K. Despite this low condensation temperature, the protosolar Hg abundance is usually assumed to be identical to the Hg abundance in CI chondrites.
Under this common, but unsupported assumption, the ratio of the terrestrial abundance of Hg divided by the CI Hg abundance yields a point that is an order of magnitude above the extrapolation of our best-fit VT in Fig. \ref{fig:TEDPentire}. 
If the protosolar Hg abundance is $13 \pm 7$ times the Hg abundance in CI chondrites, the terrestrial Hg point in Fig. 7 is lowered and agrees with our best-fit VT.
This lowering is indicated by the vertical dashed arrow beneath Hg in Fig. \ref{fig:TEDPentire}.
We propose that this agreement plausibly presents 
a better estimate of the protosolar Hg abundance than the unmodified CI chondrite Hg abundance (see footnote $l$ of Table \ref{tab:sun}). 

\subsection{Comparison of volatile depletions in the Earth and CI chondrites}
\label{sec:discu3}
\begin{figure}
	\begin{center}
		\includegraphics[trim=4cm 2cm 3cm 3cm, scale=0.6,angle=90]{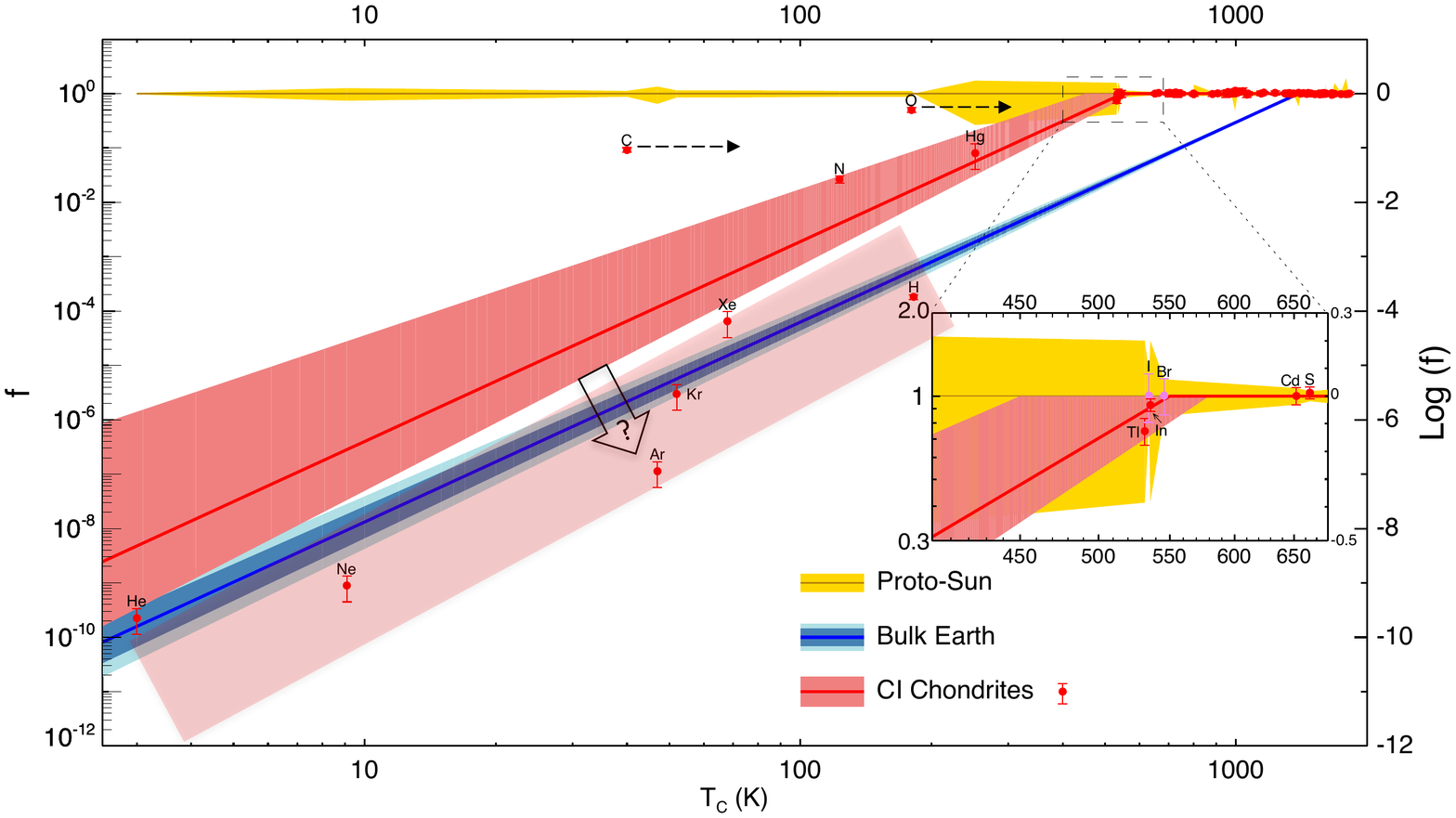} 
		\caption{The depletion of CI chondrites relative to the Sun. The thin blue diagonal wedge represents the Earth and is taken from the previous
			figure. The protosolar Hg abundance has been elevated by a factor of $13 \pm 7$ as indicated in the previous figure.
			Protosolar abundances of Tl and In are not used here. Rather their photospheric-based abundances (yellow region) and their meteoritic abundances  (red points) are plotted separately.
			Protosolar abundances of I and Br are colored in light red since they are identical to meteoritic abundances and therefore cannot help distinguish solar from meteoritic abundances.	
			The notional volatility trend (red diagonal wedge) for CI chondrites comes from assuming the same slope as the Earth's volatility trend.
			The position of the light red rectangle indicates the stronger depletion of noble gases and hydrogen. 
			For C and O, the horizontal arrows (analogous to the arrows in the previous figure) indicate modifications in their $T_C$ values that would bring their abundances into agreement with the red wedge. 
			The narrow upper end of the red wedge is characterized by the devolatilization temperature of $T_{\mathrm{D}}(\mathrm{CI}) = 550^{+20}_{-100}$ K (see inset). 
			The compositional data for CI chondrities is from \cite{Palme2014a} with 50\% uncertainties assigned to abundances of noble gases.}
		\label{fig:TEDPentireCI} 		
	\end{center}
\end{figure}

\begin{figure}[!ht]	
	\begin{center}
		\includegraphics[trim=2.5cm 4.5cm 0.5cm 6cm, scale=0.8,angle=90]{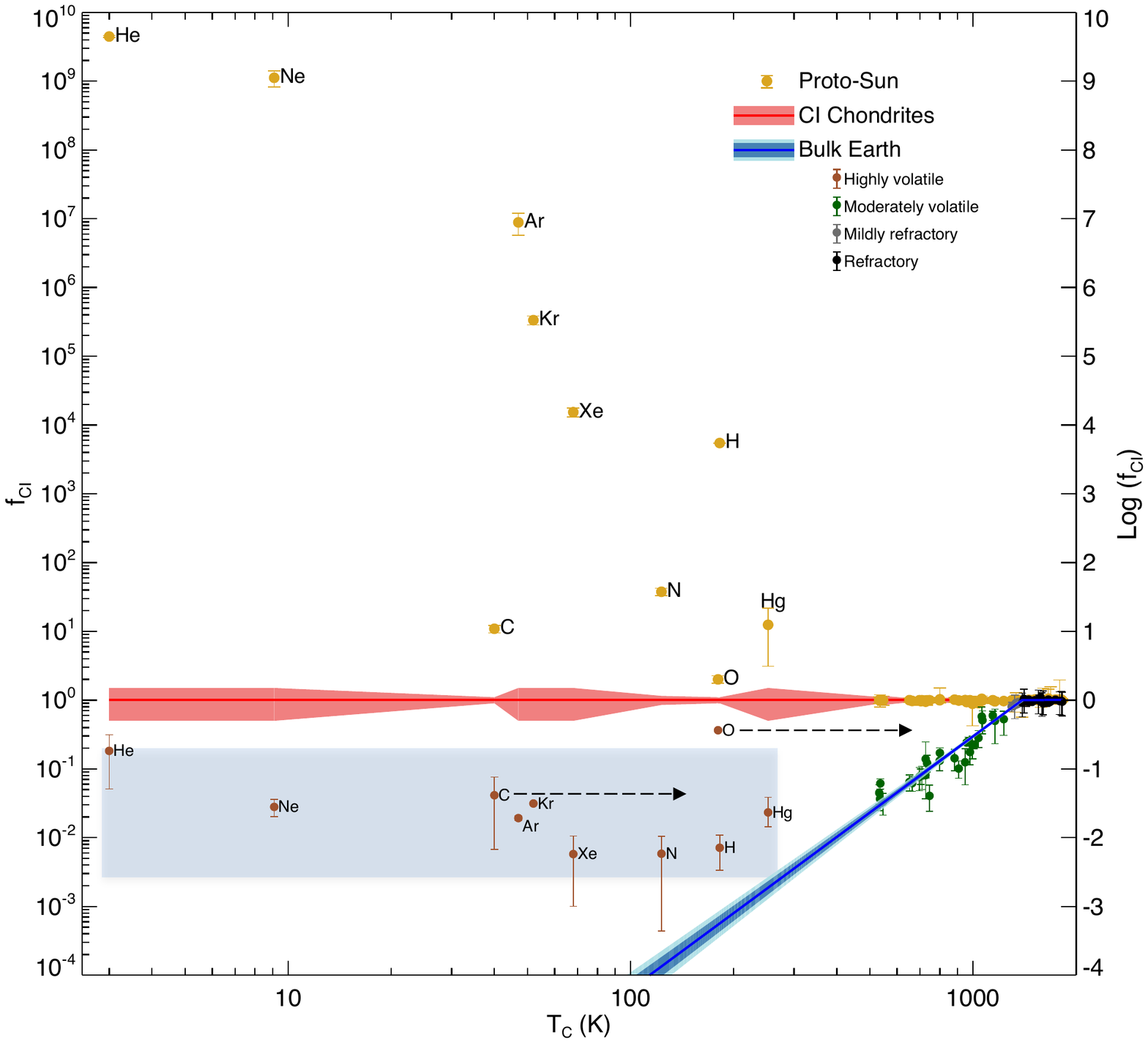} 
		\caption{Replacing the solar normalization with a normalization to CI chondrite abundances. The depletion of the highly volatile elements
			from CI chondrites makes the Sun seem enriched in these elements. The light blue rectangle again represents the position of the 
			Earth's highly volatile elements which here seem enriched compared to Earth's VT due to the depletion of these same elements
			in CI chondrites and the subsequent normalization to depleted CI chondritic values. The Hg protosolar abundance has been 
			increased by a factor of $13 \pm 7$ relative to its abundance in CI chondrites. The horizontal arrows to the right of the Earth's 
			C and O abundance are the same as those in Fig. \ref{fig:TEDPentire}.}
		\label{fig:TEDPend} 		
	\end{center}
\end{figure}

Comparing the compositions of the Earth and the Sun can help quantify the devolatilization processes involved in fractionating solar nebula material through dust grains, planetesimals and embryos, and ultimately the formation of the Earth.
Parent bodies of CI chondrites are from the outer regions of the asteroid belt and close to the snow line. They are less depleted in volatiles compared to the terrestrial planets \citep{Wasson1988}.

While terrestrial elemental abundances show depletions below $T_D\mathrm{(E)} = 1391$ K, CI chondrite abundances show depletions of highly volatile elements below $T_D\mathrm{(CI)} \approx 500$ K
(Fig. \ref{fig:TEDPentireCI}). 
As shown in the inset in Fig. \ref{fig:TEDPentireCI}, the photospheric-based solar abundances of both Tl and In are slightly higher than their corresponding CI chondritic abundances. 
This suggests that the depletion of highly volatile elements of CI chondrites (relative to the Sun) may start at the $T_C$ values of these two elements.
Without the availability of photospheric abundances of I and Br, their protosolar abundances are assumed equal to meteoritic abundances. To a first-order, therefore, the group of elements (In, Tl, I and Br) may define the boundary  $T_D(\mathrm{CI})$ between the depletion and non-depletion of the parent body of CI chondrites, analogous to the group of mildly refractory elements (e.g., Mg, Si, Fe, and Ni) which define $T_D(\mathrm{E})$ for the Earth (see inset in Fig. \ref{fig:TEDPentireCI}). Based on this data, we make a preliminary estimate and suggest that 
$T_D\mathrm{(CI)} = 550^{+20}_{-100}$ K. 
Below $T_D$(CI), the abundances of highly volatiles are depleted.
It is difficult to determine a quantitative fit to the depletion pattern of the abundances of CI chondrites because of the large scatter in these elements. However, the red wedge
in Fig. \ref{fig:TEDPentireCI} represents 
a notional CI volatility trend 
having the same slope as the terrestrial VT. 
The scatter of the highly volatile elements around this slope is very similar to that found in the Earth analysis. H and the noble gases are systematically 
depleted relative to the VT(CI).  Carbon and oxygen lie above the VT(CI) similar to the case for Earth. 
In Fig. \mbox{\ref{fig:TEDPend}} protosolar abundances are normalized to CI chondrite abundances.  
Solar abundances of the most volatile elements are systematically enriched, and terrestrial abundances of the same elements are systematically depleted.
This systemic behavior suggests that the cosmochemical abundances are controlled in the same way even for the most volatile elements.

\subsection{Beyond the quantification}
\label{sec:discu4}

Comparison of the bulk elemental compositions of Sun and Earth shows that devolatilization is an inherent process in the ultimate composition of the inner solar system. 
Comparisons of isotopic compositions of planetary materials \citep[e.g.][]{Halliday2001, Poitrasson2004, Young2009, Pringle2014, Dauphas2015, Sossi2016, Hin2017, Dauphas2017} offer complementary insights into the devolatilization processes in the early Solar System. 

While determining the specific regime responsible for a thermal gradient is beyond the scope of this paper, these conditions all but require some form of radial gradient regarding temperature in the early solar nebula.  While CAI formation is typically envisaged as occurring close to the Sun, a number of models have preferred localized thermal events such as shock fronts, lightning or planetary collisions as being responsible for these objects.  While location is important for understanding thermal processing, so too are temporal constraints.  Precision chronometry makes it increasingly plausible that chondrule formation is possible at the time of CAI formation. Thus both the radial thermal distribution of the solar nebula and more localized processing events can play a part. Modeling the physics and chemistry of devolatilization in the inner regions of protoplanetary disks is an active field of research \citep{Albarede2009, Dullemond2010, Salmeron2012a, Salmeron2012b, Wang2016, Brasser2017, Hin2017, Norris2017, Jin2018}, ultimately with the goal of providing an appropriate context for the processes active in the formation of planetary systems.

\cite{Melendez2009} and \cite{Ramirez2009} have hypothesized that the Sun could be depleted (compared to solar twins) in refractory elements because the Sun, having planets, might not have accreted as many refractories as other stars that do not have planets. Subsequent research to test this hypothesis includes \cite{Ramirez2010}, \cite{Chambers2010}, \cite{Gonzalez2014}, \cite{Liu2016}, \cite{Spina2016}, and \cite{Jermyn2018}. However, recent exoplanet statistics are consistent with all stars having some kind of planetary system in orbit around them \citep{Bovaird2015, Foreman-Mackey2016, Mulders2018}. More evidence is needed to confirm or invalidate this hypothesis.
 
\section{Summary and Conclusions}
\label{conclu}

Our main results can be summarized as follows:

\begin{enumerate}
	\item We obtain improved protosolar elemental abundances by an updated combination of current estimates of CI chondritic abundances and photospheric abundances. These new estimations indicate CI chondrites and solar abundances are consistent for 60 elements. The remaining elements either have no photospheric data (13 elements), or they are substantially depleted in CI chondrites (9 elements) or otherwise depleted in Sun (i.e. Li). 
	 
	\item Our estimate of the \textquotedblleft metallicity" (i.e. mass fraction of metals, $Z$) in the proto-Sun is 1.40\%. This value is consistent with a value of $Z$ that has been decreasing over the past three decades from 1.89\% to the current 1.40\%.

\item We have renormalized chondritic abundances ratios to the most refractory of the major elements, Al.  This results in an internally consistent normalization with all elements having abundance ratios at or below solar.  This normalization removes the apparent enrichment of refractory lithophiles produced by normalization to Mg or Si.
	
	\item The best-fit of the Earth/Sun abundance ratios $f$, to the joint devolatilization  model ($\log(f) = \alpha \log(T_C) + \beta$ and $\log(f) = 0$),
	yields $\alpha = 3.676\pm 0.142$ and $\beta = -11.556\pm 0.436$. These coefficients determine the critical devolatilization temperature 
	$T_{\mathrm{D}}(\mathrm{E}) = 1391\pm 15$ K. 
	
	\item Mercury (Hg) is the most volatile of all the elements for which the solar value has been assumed to be equal to the CI abundance. This leads to an
	expected deviation of the Earth's Hg abundance relative to the Sun. This deviation is plausibly corrected by an increase of the Sun's 
	Hg abundance by a factor of $13 \pm 7$ relative to the CI chondritic Hg abundance. 
	
	\item We use the best-fit volatility trend to derive the fractional distribution of carbon and oxygen between volatile and refractory components ($f_{\mathrm{vol}}$, $f_{\mathrm{ref}}$). The results are ($0.91\pm 0.08$, $0.09 \pm 0.08$) for carbon and ($0.80\pm 0.04$, $0.20 \pm 0.04$) for oxygen.
	
	
	\item The devolatilization processes that produced Earth's material had a similar but reduced effect on CI chondrites. 
		Analogous to the devolatilization temperature of the Earth $T_{\mathrm{D}}(\mathrm{E}) = 1391 \pm 15$ K, we estimate the devolatization temperature of CI chondrites $T_{\mathrm{D}}(\mathrm{CI}) = 550^{+20}_{-100}$ K.
\end{enumerate}

\section*{Acknowledgments}
We gratefully thank Herbert Palme and Francis Albar\`ede for their constructive reviews that improved the quality of this paper. We also thank Paolo Sossi, Martin Asplund and Thomas Nordlander for helpful comments on an earlier version of this paper. We thank David Yong for advice on calculating solar mass fractions and Katharina Lodders for information about the determination of 50\% condensation temperatures for carbon and oxygen. H.S.W. was supported by the Prime Minister's Australia Asia Endeavour Award (No. PMPGI-DEC-4014-2014) from Australian Government Department of Education and Training.

\section*{}
\bibliographystyle{elsarticle-harv} 

\appendix

\setcounter{table}{0}
\setcounter{figure}{0}
\section{How we normalize and combine photospheric and meteoritic abundances}
\label{sec:app1}
Let $N(X)$ be the photospheric abundance (by number of atoms) of element $X$. We normalize this abundance to $10^{12}$ atoms of hydrogen and write the abundance in dex as:
\begin{equation}
\label{eq:A1}
A(\mathrm{X}) |_{A(\mathrm{H})\equiv12} =\log[N(\mathrm{X})/N(\mathrm{H})] + 12.
\end{equation} 
Meteoritic abundances are often reported in parts-per-million (ppm) by mass and are denoted here as $\mathrm{X_{ppm}}$.
To combine the two sets of abundances, we normalize both to $10^{6}$ atoms of Si (i.e., $N(\mathrm{X})|_{N(Si)=10^6}$). 
As discussed in Section \ref{sec:solarabu2},
we make diffusion corrections on photospheric abundances before normalizing them and before combining them with meteoritic data. 
Following \citep{Asplund2009}, we use the overall diffusion correction factors from Turcotte \& Wimmer-Schweingruber (2002). 
Our diffusion correction converts photospheric abundances $A(\mathrm{X})_p$, into bulk solar abundances $A(\mathrm{X})_{p,0}$.
For helium we set,
\begin{equation}
\label{eq:A2}
A(\mathrm{He})_{p,0} = A(\mathrm{He})_{p} + 0.05
\end{equation} 
and for all elements heavier than helium we set,
\begin{equation}
\label{eq:A3}
A(\mathrm{X})_{p,0} = A(\mathrm{X})_{p} + 0.04.
\end{equation} 
The uncertainty in the overall diffusion correction is assumed to be 0.01 dex \citep{Asplund2009}. In Table \mbox{\ref{tab:sun}}, this 0.01 uncertainty has been added in quadrature to the uncertainties of $A(\mathrm{X})_p$ (column 3) to produce the uncertainties of $A(\mathrm{X})_{p,0}$ (column 5). 

After these diffusion corrections, photospheric abundances are normalized to Si using:
\begin{equation}
\label{eq:A4}
N(\mathrm{X})_{p,0} = 10^{A(\mathrm{X})_{p,0}-A(\mathrm{Si})_{p,0}}\times10^6
\end{equation} 
We convert the uncertainty $\sigma_{A}$ (in dex) of $A(\mathrm{X})_{p,0}$ to the standard deviation, $\sigma_{N,p,0}$ (in \%) of $N(\mathrm{X})_{p,0}$  by: 
\begin{equation}
\pm \sigma_{N,p,0} (\%) = (10^{\pm \sigma_{A}}-1)\times100
\label{eq:A5}
\end{equation} 
where the $\pm$ indicates separate upper and lower error bars.
In using Eq. \ref{eq:A5}, symmetric errors in logarithmic abundances correspond to asymmetric errors in linear abundances, and vice versa. 
Asymmetric errors add an extra level of complication to computing weighted averages.   Following \cite{Lodders2003} and \cite{Lodders2009}, 
when faced with upper and lower error bars of different sizes, we conservatively choose the larger one for further analysis.
That is why there are no separate upper and lower error bars in Table \ref{tab:sun}, column 13.


To convert meteoritic abundances by mass (i.e., $\mathrm{X_{ppm}}$) to abundances by number, normalized to Si we use:
\begin{equation}
\label{eq:A6}
N(\mathrm{X})_{m} =\frac{\mathrm{X_{ppm}}/m_a(\textup{X})}{\mathrm{Si_{ppm}}/m_a(\textup{Si})}\times10^6
\end{equation} 
where $m_a (\mathrm{X})$ is the atomic mass of an element X \citep{Wieser2013}. 
Let $\sigma_{X}$ be the uncertainty (in \%) on $\mathrm{X_{ppm}}$.
We obtain the uncertainty $\sigma_{m}$ on $N(\mathrm{X})_m$ from:
\begin{equation}
\label{eq:A7}
\sigma_{m} = N(\textup{X})_m \times \frac{\sigma_{X}}{100}
\end{equation}  
With these conversions, the two data sets are ready to be combined to obtain protosolar abundances.
For the 60 elements in Table \ref{tab:sm} for which we compute the weighted average of the two sets, we use
\begin{equation}
\label{eq:A8}
N(\mathrm{X})_{0} =\frac{N(\mathrm{X})_{p,0}/\sigma_{N,p,0}^2 + N(\mathrm{X})_{m}/\sigma_{m}^2}{1/\sigma_{N,p,0}^2 + 1/\sigma_{m}^2}
\end{equation} 
If the values of $N(\mathrm{X})_{p,0}$ and $N(\mathrm{X})_{m}$ overlap within uncertainties 
we compute the uncertainty on $N(\mathrm{X})_{0}$ using
\begin{equation}
\label{eq:A9}
\sigma_{N_{0}} = \sqrt{1/(1/\sigma_{N,p,0}^2 + 1/\sigma_{m}^2)}
\end{equation}
If the values of $N(\mathrm{X})_{p,0}$ and $N(\mathrm{X})_{m}$ do not overlap within uncertainties
the highest upper and lowest lower limit of the two data points are taken as the error bars.  

\begin{table*}
	{\scriptsize
		\centering
		\caption{Combining photospheric and meteoritic abundances: how our method compares with the method of \cite{Lodders2009}}
		\begin{tabular}{c p{6cm} c || p{6cm} c}
			\toprule
			\multicolumn{1}{c}{SM$^\#$} &  \multicolumn{2}{c}{This Work} & \multicolumn{2}{c}{\cite{Lodders2009}}\\
			& Elements & Number & Elements & Number\\
			\hline  
			p   &   H, C, N, O, He, Ne, Ar, Kr, Xe   &  9  & H, C, N, O, He, Ne, Ar, Kr, Xe  & 9\\
			& & & & \\
			m  &   Li, As, Se, Br, Sb, Te, I, Cs, Ta, Re, Pt, Hg$^\dagger$, Bi, U   & 14 &  Li, As, Se, Br, Sb, Te, I, Cs, Ta, Re, Pt, Hg, Bi, U, Be, B, F, Si, Cl, Ca, c, Ti, Mn, Cu, Zn, Ga, Rb, Sr, Ru, Ag, Cd, In, Sn, La, Nd, Sm, Tb, Ho, Tm, Yb, Lu, Hf, W, Os, Au, Tl, Pb, Th & 48  \\
			& & & & \\
			&  &  &  &  \\
			a& Na, Mg$^*$, Al, P, S, K, V, Cr, Fe, Co, Ni, Ge, Y, Zr, Nb, Mo, Rh$^*$, Pd$^*$, Ba, Ce, Pr, Eu, Gd, Dy, Er, Ir$^*$, Ru, Cd, In, Sn, Nd, Sm, Tb, Ho,  Be, B, F, Si, Cl, Ca, Mn, Ti, Tm, Yb, Cu, Zn, Ga, Sr, Lu, Os, Au, Tl, Pb, Th, Sc$^*$, Ag$^*$, La$^*$, Hf$^*$, Rb$^*$, W$^*$ &  60  & Na, Mg, Al, P, S, K, V, Cr, Fe, Co, Ni, Ge, Y, Zr, Nb, Mo, Rh, Pd, Ba, Ce, Pr, Eu, Gd, Dy, Er, Ir & 26\\
			
			
			\bottomrule
			\\
			\multicolumn{5}{p{16cm}}{\scriptsize $\#$ SM- Selection methods: \textquoteleft p'- photospheric data only; \textquoteleft m'- meteoritic data only; \textquoteleft 'a'- weighted average of photospheric and meteoritic data} \\
			\multicolumn{5}{p{16cm}}{\scriptsize $^\dagger$ In Section \ref{sec:discu2c}, we propose that protosolar Hg abundance should be elevated by a factor of $13\pm7$ over CI chondritic Hg abundance.}\\
			\multicolumn{5}{p{16cm}}{\scriptsize $^*$ For each of these 10 elements, uncertainty in the meteoritic abundance does not overlap with the uncertainty in the photospheric abundance. We compute the weighted average in the usual way (Eq. \ref{eq:A8}),
				however to be conservative, for the uncertainties, instead of using Eq. \ref{eq:A9}, 
				we use the upper and lower limits of the uncertainties of the photospheric and meteoritic abundances. 
			} \\
		\end{tabular}
		\label{tab:sm}
	}
\end{table*}

\begin{figure}[!htbp]	
	
	\begin{center}
		\includegraphics[trim=1.0cm 0.8cm 1.2cm 1.4cm, scale=0.6,angle=90]{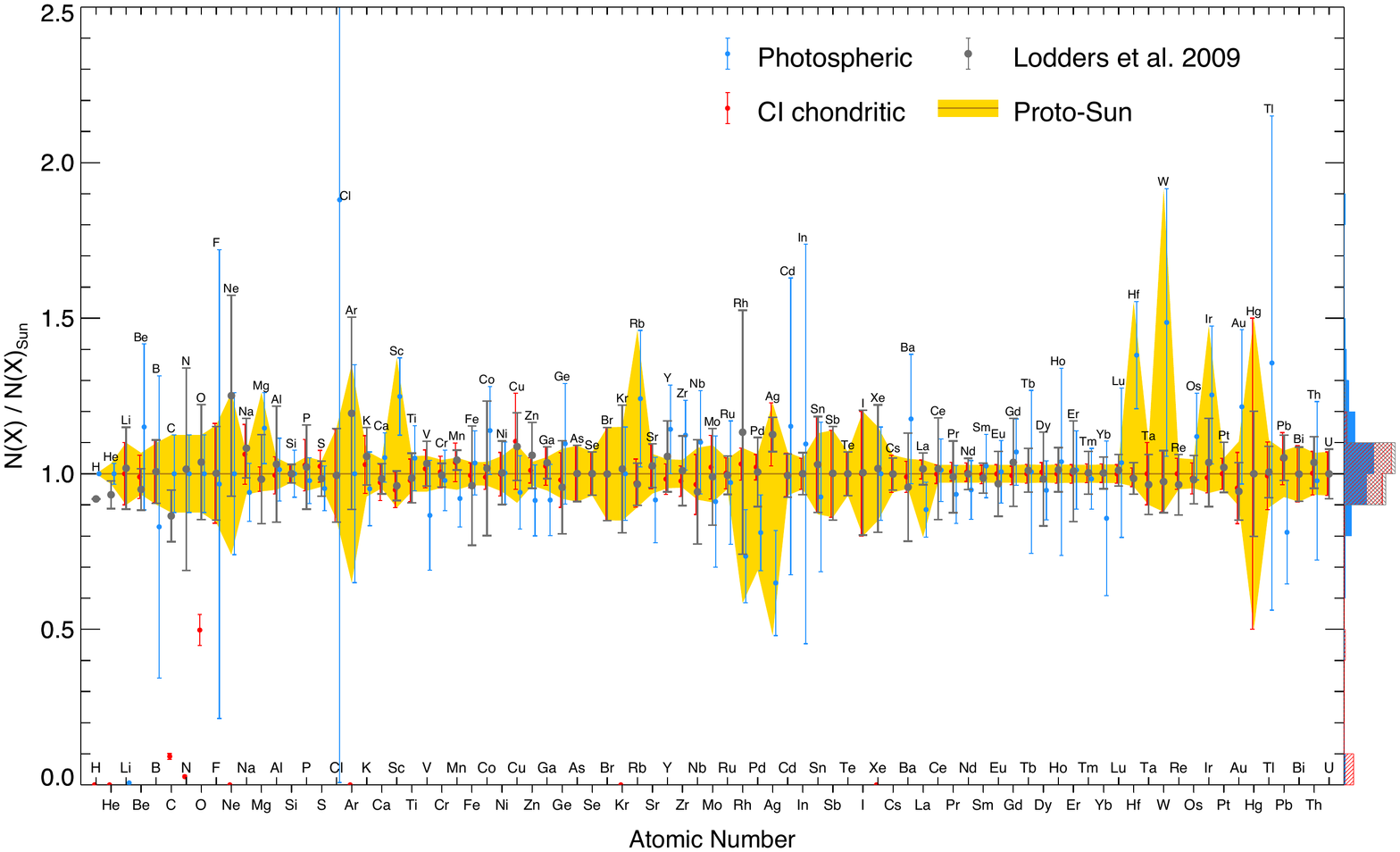} 
		\caption{Same as the lower panel of Fig. \ref{fig:sun-2}, but with the \cite{Lodders2009} estimates for protosolar elemental abundances (dots in grey) over-plotted. Overall, the \cite{Lodders2009} estimates are consistent with CI chondritic abundances as shown in the comparison between grey and red dots and in the grey and red histograms along the right side of the plot. That is because \cite{Lodders2009} has placed 48 elements in the second (meteorite-only) group -- compared to 14 in this work -- as shown in Table \ref{tab:sm}. For elements with inconsistent abundances between meteoritic and photospheric data, like Sc, Rb, Pd, Ag, Hf, W and Ir, our estimates take both the meteoritic and solar photospheric data into account while \cite{Lodders2009} uses only meteoritic data. This comparison is based on a normalization to Si by number and without considering the effects of radioactive decay and fusion on both estimates. The reasons the H and He abundances in \cite{Lodders2009} are lower than our estimates include i) the downward estimate of Si photospheric abundance in our compilation (compared to the compilation of \cite{Lodders2009}) and ii) larger diffusion corrections adopted in \cite{Lodders2003} (0.061 for He and 0.053 for all elements heavier than He) compared to our corrections (see Equations \ref{eq:A2} and \ref{eq:A3}). The improvement in the estimates of protosolar abundances of C, N, O and noble gases is due to our compilation of photospheric abundances coming from the most recent literature as described in Section \ref{sec:solarabu1}. 
	    }
		\label{fig:suncompare} 		
	\end{center}
\end{figure}

\begin{table*}[!htbp]
	{\scriptsize
		\centering
		\caption{Protosolar mass fractions of H ($X$), He ($Y$) and metals ($Z$).} 
		\begin{center}
			\begin{tabular}{l  c c c c}
				\toprule
				Source & $X$ & $Y$ & $Z$ & $Z/X$ \\
				\cmidrule{1-5}  
				\cite{Anders1989} & 0.7068 $\pm$ 0.0177  & 0.2743 $\pm$ 0.0165 & 0.0189 $\pm$ 0.0016& 0.0267 $\pm$ 0.0024\\
				\cite{Grevesse1998} & 0.7090 $\pm$ 0.0100 & 0.2750 $\pm$ 0.0100 & 0.0160 $\pm$ 0.0016 & 0.0230 $\pm$ 0.0023\\
				\cite{Lodders2003} & 0.7110 $\pm$ 0.0040 & 0.2741 $\pm$ 0.0120 & 0.0149 $\pm$ 0.0015 & 0.0210 $\pm$ 0.0021\\
				\cite{Lodders2009} & 0.7112 $\pm$ 0.0033 & 0.2735 $\pm$ 0.0036 & 0.0153 $\pm$ 0.0014 & 0.0215 $\pm$ 0.0019\\
				\cite{Asplund2009} & 0.7154 $\pm$ 0.0037 & 0.2703 $\pm$ 0.0037& 0.0142 $\pm$ 0.0009 & 0.0199 $\pm$ 0.0013\\
				A09-S15-G15$^{a}$ & 0.7156 $\pm$ 0.0037 & 0.2703 $\pm$ 0.0037 & 0.0141 $\pm$ 0.0009 & 0.0197 $\pm$ 0.0013\\
				\bf This work & \bf 0.7157 $\pm$ 0.0037 &\bf 0.2703 $\pm$ 0.0037 &\bf 0.0140 $\pm$ 0.0009 &\bf 0.0195 $\pm$ 0.0013\\
				\bottomrule
				\\
				\multicolumn{5}{p{12cm}}{\textit{Note:} Mass fractions not presented in specific references are computed from the corresponding estimate of protosolar abundances in the reference and constrained by the corresponding H or He mass fraction from helioseismology used in the reference. Uncertainties not presented in a reference, associated with the mass fractions, are simulated by a standard Monte Carlo approach.}\\ 
				\multicolumn{5}{p{12cm}}{\scriptsize$^a$ A09-S15-G15 is the combination of \cite{Asplund2009}, \cite{Scott2015a, Scott2015b} and \cite{Grevesse2015}}\\				
			\end{tabular}
			\label{tab:solarfrac}
		\end{center}
	}
\end{table*}

\setcounter{table}{0}
\setcounter{figure}{0}

\section{How we renormalize meteoritic abundances from a silicon to a hydrogen normalization}
\label{sec:app2}
\begin{figure*}[!htbp]
	\begin{center}
		\includegraphics[trim=5cm 1cm 5cm 2cm, scale=0.6,angle=90]{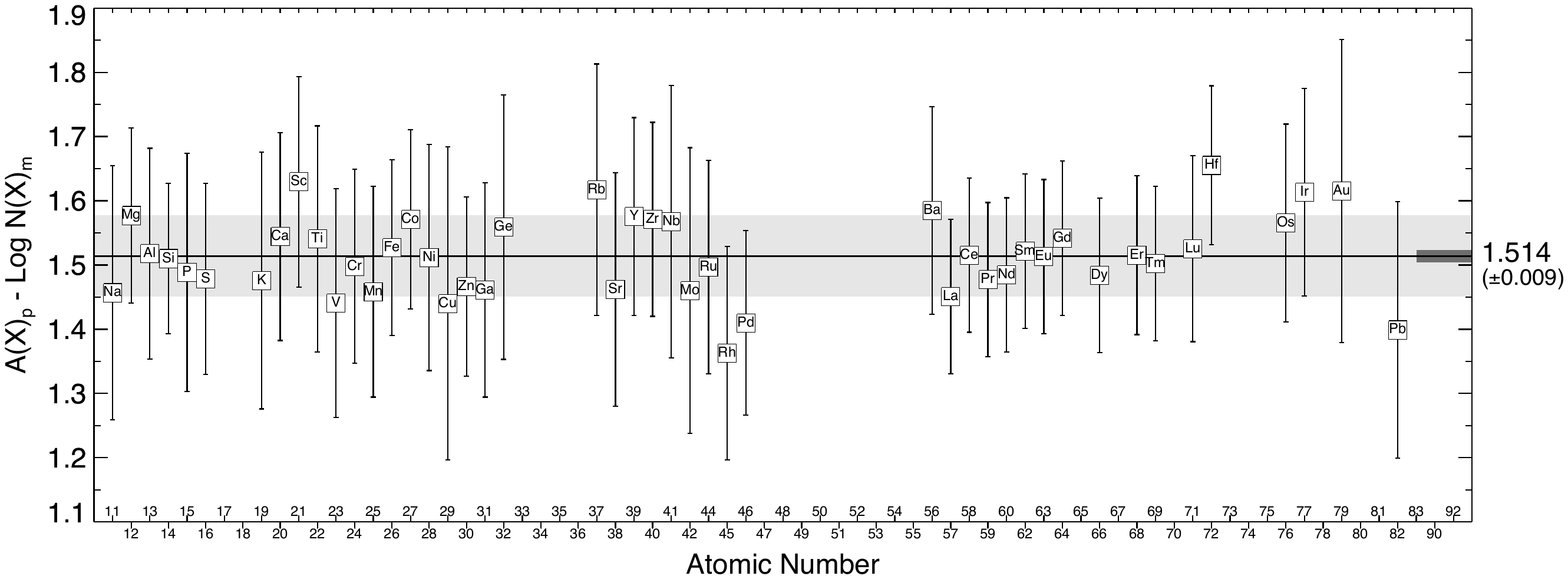}  
		\caption{Values of $\Delta$ from Eq. \ref{eq:B3} for 46 refractory elements. The gray band shows the 1$\sigma$ standard deviation. On the far right we plot our main result: $\Delta = 1.514 \pm 0.009$.} 
		\label{fig:lin2log}
	\end{center}
\end{figure*}

Due to the depletion of hydrogen in meteorites, Eq. \ref{eq:A1} is not used to convert Si-normalized meteoritic abundances to H-normalized abundances.
Instead we write, 

\begin{equation}
\label{eq:B1}
A(\mathrm{X})_m = \log N(\mathrm{X})_m + \Delta
\end{equation}
where $\Delta$ is a numerical conversion factor.
For refractory elements, solar abundances and CI abundances are indistinguishable. Therefore we expect,
\begin{equation}
\label{eq:B2}
A(\mathrm{X})_m = A(\mathrm{X})_p
\end{equation}
Combining Eqs. \ref{eq:B1} and \ref{eq:B2} we obtain
\begin{equation}
\label{eq:B3}
\Delta = A(\mathrm{X})_p - \log N(\mathrm{X})_m
\end{equation}
Following \cite{Lodders2009} we require that the elements used in Eqs \ref{eq:B1} - \ref{eq:B3}, satisfy several criteria:
(1) the elements must have both photospheric and meteoritic abundances,
(2) uncertainties on their photospheric abundances must be less than 0.1 dex (i.e., below $\sim$ 25\%),
(3) no noble gases,
(4) atomic numbers greater than neon.
Thus, for these elements, by subtracting the logarithm of the Si-normalized meteoritic abundances ($\log N(\textup{X})_m$) from the logarithmic H-normalized photospheric abundances  ($A(\textup{X})_p$) we obtain an average numerical factor $\Delta$ from Eq. \ref{eq:B3} that can be generallly used in Eq. \ref{eq:B1}
to convert Si-normalized meteoritic abundances to H-normalized meteoritic abundances.
Using the updated meteoritic and photospheric abundances compiled here, 46 elements satisfy the criteria above.
We find $\Delta = 1.514 \pm 0.009$ where the uncertainty is the standard error of the mean:  
$\frac{1}{\sqrt{n}} \: \sum_{i=1}^{n=46} \sqrt{(x_i-\bar{x})^2/(n-1)}$.
Fig. \ref{fig:lin2log} shows the various values of $\Delta$ from Eq. \ref{eq:B3} as a function of atomic number. 
Over the past few decades, estimates of $\Delta$ have been decreasing.  
This decreaase can be attributed to the updates of meteoritic and photospheric data and an increasing number of elements
that can be used. For example,
\cite{Anders1989} used 12 elements and obtained 1.554.
\cite{Lodders2003} used 35 elements and obtained 1.540.
\cite{Lodders2009} used 39 elements and obtained  1.533.
We used 46 elements and obtained $1.514 \pm 0.009$, significantly lower than previous estimates.

When meteoritic abundances are taken as a proxy for protosolar abundances, a diffusion correction factor should be added to $\Delta$ to make the meteoritic abundances comparable with protosolar abundances. 
Because the diffusion correction factor we applied for all elements (except helium) is 0.04 dex, the corrected conversion factor should be $1.554 \: (\: = 1.514 + 0.04)$  when taking photospheric diffusion into account. 

\setcounter{table}{0}
\setcounter{figure}{0}
\section{$\chi^2$ minimization to determine the coefficients of the volatility trend}
\label{sec:app4}
\begin{figure*}[!htbp]
	\begin{center}
		\includegraphics[trim=0.5cm 5.5cm 1cm 6cm, scale=0.8,angle=0]{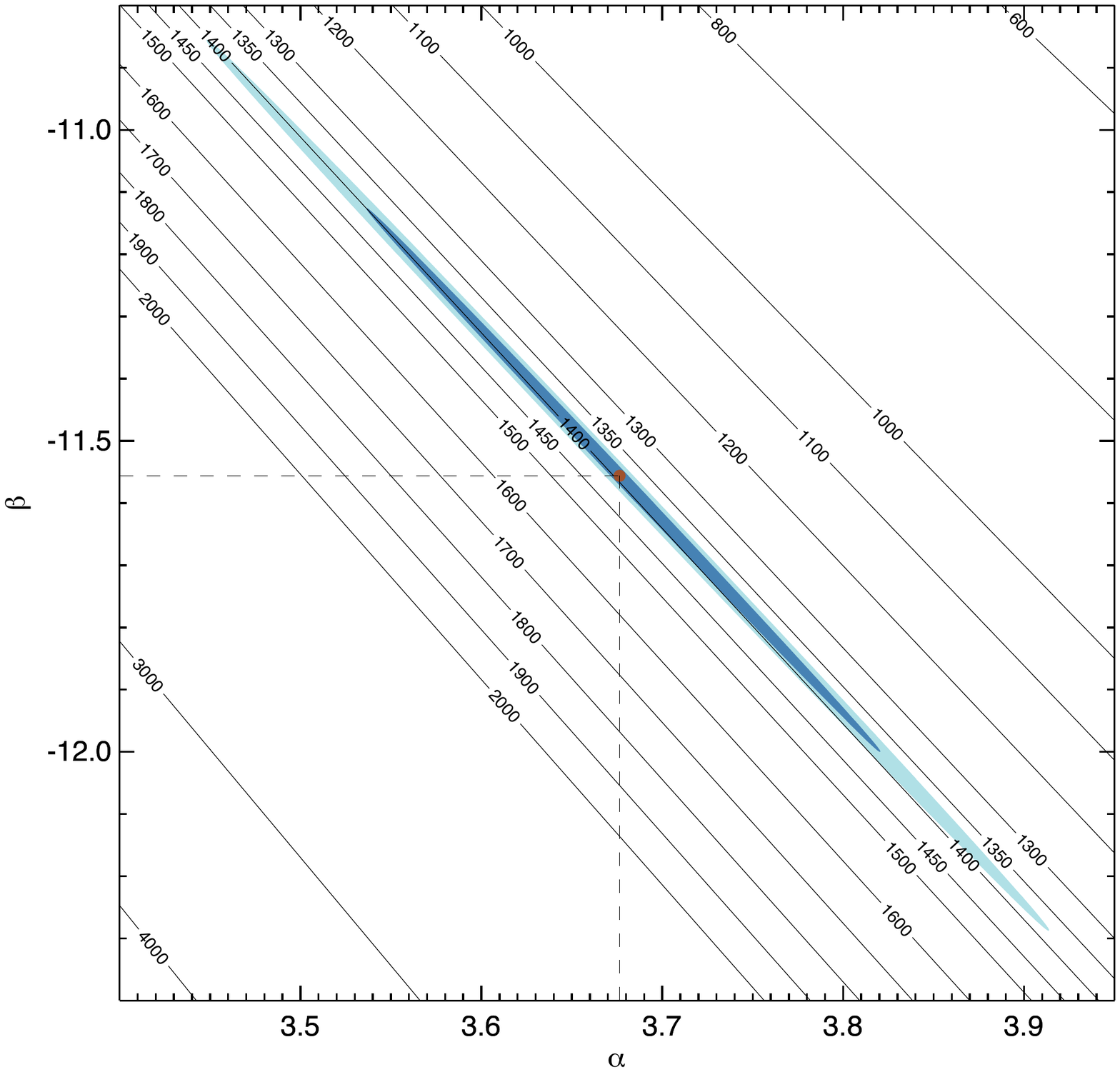}  
		\caption{Determination of the coefficients $\alpha$ and $\beta$ and therefore of $T_{\mathrm{D}}(\mathrm{E}) = 10^{-\beta/\alpha}$.
			The central brown dot represents the best fit values $\alpha =3.676$ and $\beta = -11.556$ indicated by the vertical and horizontal dash lines. 
			The dark and light blue ellipses indicate respectively, the 68\% and 95\% confidence levels. These levels are computed from
			$\chi_{\mathrm{min}}^2 +2.3$ and $\chi_{\mathrm{min}}^2 + 6.17$ respectively.
			The diagonal solid black lines labeled between 600 and 4000 are isothermal lines of $T_D$ (=$10^{-\beta/\alpha}$) in units of Kelvin.
			The best fit devolatilization temperature $T_{\mathrm{D}}(\mathrm{E}) = 1391 \pm 15$ K.} 
		\label{fig:chisqarr}
	\end{center}
\end{figure*}

The elemental abundance ratios of bulk Earth to the proto-Sun normalized to Al are $ f = N\mathrm(X)_{\mathrm{Earth}} / N\mathrm(X)_{\mathrm{Sun}}$. 
where $N\mathrm(X)_{\mathrm{Sun}}$ is from column 13 of Table \ref{tab:sun} after the Al normalization,
and $N\mathrm(X)_{\mathrm{Earth}}$ is from  Wang et al 2017. 
These ratios are plotted as a function of $T_C$ in Fig. 5, where
$T_C$ is the elemental 50\% condensation temperature for the relevant element \citep{Lodders2003}.  We fit all elements
with $T_C > 500$ K. 
We simultaneously fit the two functions $\log(f) = \alpha\log(T_C) + \beta$ and $\log(f) = 0$ to the abundance data, 
and find the best-fit values of  $\alpha$ and $\beta$.   

\textbf{Step 1}: We create an array of the two parameters $\alpha$ and $\beta$ in the ranges $2 < \alpha < 6$ and
$-20 < \beta < -6$, using an increment step of 0.001 for both parameters.
For each pair $(\alpha,\beta)$, we compute $T_{\mathrm{D}}(\alpha,\beta) = 10^{-\beta/\alpha}$. 
For each coordinate pair $(\alpha,\beta)$, elements with $T_C < T_{\mathrm{D}}(\alpha,\beta)$ are fit to $\log(f) = \alpha\log(T_C) + \beta$ 
while elements with $T_C > T_{\mathrm{D}}(\alpha,\beta)$ are fit to $\log(f) = 0$. 

\textbf{Step 2}: We compute $\chi^2 = \chi^2_\mathrm{I} + \chi^2_\mathrm{II}$ using the equations: 
\begin{equation}
\label{eq:D1}
\chi^2_\mathrm{I}(\alpha, \beta) = \sum_{i=1}^{k}\frac{[log (f_i) - (\alpha\log(T_{C_i}) + \beta)]^2}{\sigma_i^2} 
\end{equation} 
where $k$ is the number of the last element whose condensation temperature is less than $T_{\mathrm{D}}(\alpha,\beta)$. 
Thus $k$ is a function of $(\alpha,\beta)$.
For the remainder of the elements, $T_C > T_{\mathrm{D}}(\alpha,\beta)$ and we compute,

\begin{equation}
\label{eq:D2}
\chi^2_{\mathrm{II}}(\alpha, \beta) = \sum_{i=k+1}^{N}\frac{[log (f_i) - 0]^2}{\sigma_i^2} 
\end{equation} 
where $N =72$. This comes from  83 (the total number of elements) minus 11 ( number of elements with $T_C < 500$ K).
In both equations, when the error bars of an elemental abundance are asymmetric, the error bar in the direction of the model is used for $\sigma_i$.

We find $\chi_{\mathrm{min}}^2 = 52.67$, which is the sum of $\chi_{\mathrm{I}, min}^2 = 43.68$ and $\chi_{\mathrm{II}, min}^2 = 8.99$. The best fit coefficient values are $\alpha = 3.676 \pm 0.142$ and $\beta = -11.556 \pm 0.436$. Therefore, the best fit devolatilization temperature $T_\mathrm{D}$(E) = 1391 $\pm$ 15 since $T_\mathrm{D}=10^{-\beta/\alpha}$. 
The number of elements with $T_C < 1391$ K is 37.  Thus the approximate number of degrees of freedom for the first part of the fit (Eq. \ref{eq:D1})
is $\nu_{I} =35 ( = 37 - 2)$. For the second part of the fit (Eq. \ref{eq:D2}), the number of degrees of freedom is not equal to the number of data points, since protosolar and terrestrial abundances of refractory elements are highly correlated. We approximate the number of degrees of freedom for the second part as  
$\nu_{II} \approx \chi_{\mathrm{II}}^2 = 9$. Thus, with $\nu = \nu_{I} + \nu_{II}$ the reduced $\chi^2$ of the best fit is $\chi_{\mathrm{min}}^2/\nu = 52.67/44 \approx 1.2$. The probability of having a reduced $\chi^2$ lower than this value is $P( \chi^2 \le \chi_{\mathrm{min}}^{2}, \nu)=0.83$.  Thus, our best-fit is a reasonably good fit.

The coefficients and $T_{\mathrm{D}}(\mathrm{E})$ of our quantified VT are listed in Table \ref{tab:patternfunc}, along with those of previous analogous VTs.
\begin{table*}[!htbp]
	{\small
		\centering
		\caption{Comparison of our analysis with previous studies with respect to the coefficients $\alpha$ and $\beta$ and critical devolatilization temperatures $T_{\mathrm{D}}(\mathrm{E})$}
		\begin{center}
			\begin{tabular}{l  l l  l }
				\toprule
				\multirow{2}{1.0cm}{Source} & \multicolumn{2}{c}{Coefficients} & \multirow{2}{1.5cm}{$T_{\mathrm{D}}$ (E)$^a$ [K]} \\ 
				\cmidrule{2-3}
				& $\alpha$ & $\beta$ &  \\ 
				\hline  
				\cite{Kargel1993}$^b$  & 3.246 & -9.792  & 1427 \\
				&       &  (-10.239) & \\[2pt]
				\cite{McDonough2014}$^c$ & 3.78 & -11.82 & 1415 \\ 
				&       &  (-11.91) & \\[2pt]
				\cite{Palme2014b}$^d$ & 2.66 to 4.27 & -8.23 to -13.30 & 1356  \\  
				&       &  (-8.32 to -13.38) & \\[2pt]
				\bf This work & \bf 3.676 $\pm$ 0.142 & \bf -11.556 $\pm$ 0.436 & \bf 1391 $\pm$ 15 \\ 
				\bottomrule
				\multicolumn{4}{p{10cm}}{\scriptsize $^a$ $T_{\mathrm{D}}$ is calculated using renormalized coefficients through $10^{-\beta/\alpha}$, except for \cite{Palme2014b}, $T_D$ of which is fixed at the $T_C$ of Eu.}	\\
				\multicolumn{4}{p{10cm}}{\scriptsize $^b$ \cite{Kargel1993} parameterized a VT as $\log(\mathrm{BSE/CI}) = 3.246\log(T_C) - 9.792$. 
					This VT has a factor of refractory lithophile enrichment (RLE) $\sim$ 2.8 in BSE relative to CI chondrites. 
					We remove RLE, by reducing the initial $\beta$ (-9.792) by $\log(2.8)$. The renormalized $\beta$ is -10.239.} \\
				\multicolumn{4}{p{10cm}}{\scriptsize $^c$ \cite{McDonough2014} normalized to Mg in CI chondrites (i.e., (X/Mg)$_{BSE}$ / (X/Mg)$_{CI}$), to produce an indicative VT in their Fig. 4. By assuming a functional form $\log(\mathrm{BSE/CI}) = \alpha\log(T_C) + \beta$, we estimate its coefficients ($\alpha \approx 3.78$ and $\beta \approx -11.82$). This VT has a factor of RLE $\sim$ 1.18. We renormalize by reducing  the initial $\beta$ by $\log(1.18)$. The renormalized $\beta$ is -11.91.} \\
				\multicolumn{4}{p{10cm}}{\scriptsize $^d$ \cite{Palme2014b} normalized to Mg in CI chondrites (i.e., (X/Mg)$_{BSE}$ / (X/Mg)$_{CI}$), to produce an indicative VT in their Fig. 21 (on a $T_C$-$\log(\mathrm{BSE/CI})$ plot). 
					Thus, a functional form $\log(\mathrm{BSE/CI}) = a T_C + b$ is assumed. 
					By fixing its $T_D$(E) at the $T_C$  of Eu (1356 K), the ranges of $a$ and $b$ are
					$[0.00134,0.00216]$ and $[-1.7367,-2.8460]$, respectively. 
					In a log-log form, the corresponding $\alpha$ and $\beta$ are in the range $[2.66, 4.27]$ and $[-8.23, -13.30]$ respectively. The corresponding VT has an RLE factor of $\sim$ 1.22. 
					We renormalize by reducing  the initial $\beta$ by $\log(1.22)$.  
					The renormalized $\beta$ is in the range $[-8.32, -13.38]$.
				} \\
				
			\end{tabular}
		\end{center}
		\label{tab:patternfunc}
	}
\end{table*}






\end{document}